\documentclass{aa}
\usepackage[varg]{txfonts}
\usepackage{graphicx,natbib,bm,gensymb}
\usepackage[colorlinks,citecolor=blue,linkcolor=blue]{hyperref}
\usepackage{csquotes}	
\usepackage{siunitx}
\usepackage{pifont}
\newcommand{\cmark}{\ding{51}}%
\newcommand{\xmark}{\ding{55}}%
\DeclareSIUnit\Mearth{M_{\oplus}}
\DeclareSIUnit\Rearth{R_{\oplus}}
\DeclareSIUnit\Msun{M_{\odot}}
\DeclareSIUnit\Rsun{R_{\odot}}
\DeclareSIUnit\Lsun{L_{\odot}}
\DeclareSIUnit\Zsol{Z_{\odot}}
\DeclareSIUnit\Mstar{M_{*}}
\DeclareSIUnit\Rstar{R_{*}}
\DeclareSIUnit\Lstar{L_{*}}
\DeclareSIUnit\year{yr}
\DeclareSIUnit\month{month}
\DeclareSIUnit\lightyear{ly}
\DeclareSIUnit\AU{au}
\DeclareSIUnit\parsec{pc}
\DeclareSIUnit\erg{erg} 				
\DeclareSIUnit\eV{eV} 	
\DeclareSIUnit\RA{RA}
\DeclareSIUnit\DEC{DEC}	
\DeclareSIUnit\dex{dex}	
\DeclareSIUnit\mag{mag}	
\graphicspath{{./}{figures/}}
\bibpunct[, ]{(}{)}{;}{a}{,}{,}

\newcommand{\revised}[1]{{#1}}

\makeatletter
\newcommand{\bianca}{\renewcommand\NAT@open{[}\renewcommand\NAT@close{]}}
\makeatother

\begin{document}

\title{Dust entrainment in magnetically and thermally driven disk winds}

\author{P. J. Rodenkirch\inst{1} \and
C. P. Dullemond \inst{1}}
\institute{Institute for Theoretical Astrophysics, Zentrum f\"ur Astronomie, Heidelberg University, Albert-Ueberle-Str. 2, 69120 Heidelberg, Germany
  }

\date{\today}
 
\abstract{
    Magnetically and thermally driven disk winds have gained popularity
    in the light of the current paradigm of low viscosities in protoplanetary disks
    that nevertheless present large accretion rates even in the presence of inner cavities.
    The possiblity of dust entrainment in these winds may explain 
    recent scattered light observations and constitutes a way of dust transport towards
    outer regions of the disk.
}
{
    We aim to study the dust dynamics in these winds and explore the 
    differences between photoevaporation and magnetically driven disk winds in this regard.
    We quantify maximum entrainable grain sizes, the flow angle, and the general
    detectability of such dusty winds.
}
{
    We used the FARGO3D code to perform global, 2.5D axisymmetric, nonideal
    MHD simulations including ohmic and ambipolar diffusion. Dust was treated 
    as a pressureless fluid. Synthetic observations were created with
    the radiative transfer code RADMC-3D.
}
{
    We find a significant difference in the dust entrainment efficiency
    of warm, ionized winds such as photoevaporation and 
    magnetic winds including X-ray and extreme ultraviolet (XEUV) heating compared to cold magnetic winds.
    The maximum entrainable grain size varies from $\SI{3}{\micro\m} - \SI{6}{\micro\m}$
    for ionized winds to $\SI{1}{\micro\m}$ for cold magnetic winds.
    The dust flow angle decreases rapidly with increasing grain size.
    Dust grains in cold magnetic winds tend to flow along a shallower angle 
    compared to the warm, ionized winds.
    With increasing distance to the central star, the dust 
    entrainment efficiency decreases. Larger values of the turbulent
    viscosity increase the maximum grain size radius of possible dust entrainment.
    Our simulations indicate that diminishing dust content in the outer regions of the wind 
    can be mainly attributed to the dust settling in the disk. The Stokes number
    along the wind lauching front stays constant in the outer region.
    In the synthetic images, the dusty wind appears as a faint, conical emission
    region which is brighter for a cold magnetic wind.
}{}


\titlerunning{Dust entrainment in disk winds}

\maketitle

\section{Introduction}
\label{sec:intro}
The limited lifetime of a few million years \citep{Haisch2001, Mamajek2004,
Ribas2015a} of protoplanetary disks has sparked interest in various mechanisms
that transport and extract angular momentum from the disk.
One possibility to drive accretion onto the central star is 
viscous turbulence, usually referred to as $\alpha$-viscosity
in the context of disks \citep{Shakura1973}.
Turbulence might be generated by a multitude of physical 
instabilities in gas and dust. The well studied 
magnetorotational instability (MRI) \citep{balbus_1991}
present in an ionized, differentially rotating disk threaded by 
a weak magnetic field can lead to values of $\alpha$ roughly ranging from 
$10^{-3}$ to $10^{-2}$ and it was studied in a
local shearing box \citep{fromang_2007a, fromang_2007b, Bai2011a}
as well as in global simulations \citep{dzyurkevich_2010, Flock2011}.\\
The MRI can only operate if the ionization fraction of gas is sufficiently 
high, which is not necessarily the case in the midplane of the disk \citep{igea_1999_ionization}, 
where nonideal magnetohydrodynamic (MHD) effects such as ohmic diffusion dominate.
These so-called dead zones prevent magnetically driven accretion 
in the midplane, whereas layered accretion in the ionized upper part 
of the disk atmosphere is still possible \citep{gammie_1996}. 
In the more dilute outer regions of the disk, the ionization fraction rises again
and the edge between the inner dead zone and the outer MRI-active part
might cause jumps in the density profile and therewith enable dust or pebble traps
\citep{lyra_2009_deadzone_edge, dzyurkevich_2010, Dzyurkevich2013b}.
Purely hydrodynamic instabilities such as the global baroclinic 
instability \citep{klahr_2003} and the convective overstability 
\citep{Klahr2014}, on the other hand, are alternative 
turbulence driving mechanisms that can also operate in the weakly ionized 
parts of the disk. In the presence of short gas cooling times, 
the vertical shear instability (VSI) causes vertical oscillations
which are able to transport small dust grains towards the upper layers of the disk
\citep{stoll_vsi_2014, stoll_vsi_2016, flock_vsi_2020, blanco_vsi_2021}.
Vortices triggered by the VSI in combination with the Rossby wave instability
\citep{lovelace_1999} may act as long-lived dust traps \citep{manger_2018,
pfeil_klahr_vsi_2021}.\\
Since recent observation hint towards small viscosity and thus
small values of $\alpha$ on the order of $10^{-4}$ to
$10^{-3}$ \citep{flaherty_2015, flaherty_2017},
the effect of disk winds might be a possible explanation for the
finite disk lifetimes.
In the context of photoevaporation, ionizing radiation either from the central star
or the from the stellar environment
heats the surface layers of the disk and drives winds due to the thermal
pressure gradient. 
Depending on the dominant type of radiation, the wind mass loss rates are on the
order of $\SI{e-10}{\Msun\per\year}$ with EUV radiation \citep{Hollenbach1994,
font_2004}, and $\SI{e-8}{\Msun\per\year}$ to $\SI{e-7}{\Msun\per\year}$ with
FUV radiation \citep{Adams2004, Gorti2009} and X-ray radiation
\citep{Ercolano2009, Owen2010, Owen2012, Picogna2019, ercolano_2021, picogna_2021}.\\
An alternative way to produce strong disk winds would be by considering the effect of magnetic fields.
In the presence of a global magnetic field the differential shearing motion of
the disk generates a magnetic pressure gradient, accelerating matter and thereby
driving the outflow. In the strong field limit, the gas follows the 
magnetic field lines and is magneto-centrifugally accelerated, as described 
in \cite{Blandford1982}. Contrary to photoevaporation, magnetically driven
winds extract angular momentum from the disk and drive accretion.
The accretion flow has to pass by the mostly vertically aligned magnetic field
in the disk which only allows stationary solutions if the coupling between gas and 
magnetic field is weak, thus if significant diffusive nonideal MHD effects
are present \citep{wardle_1993, ferreira_1993, Ferreira1995}.
The magnetic launching mechanism is considered to be responsible for high-velocity jet
outflows and has been studied extensively in global resistive simulations 
\citep{Casse2002, Zanni2007, tzeferacos_2009, Sheikhnezami2012, Fendt2013}.
\cite{stepanovs_2014} and \cite{Stepanovs2016} determined the  
transition between magnetocentrifugally driven and magnetic 
pressure-driven jets at $\beta \approx 100$ where $\beta$ is the ratio
of thermal over magnetic pressure. \\
With the growing interest in winds in the context of protoplanetary disks,
numerous studies have been performed over the last decade, first in the form 
of local shearing box simulations \citep{Suzuki2009a, suzuki_2010, fromang_2013, bai_2013a, bai_2013b}.
Global nonideal MHD simulations have been carried out by \cite{Gressel2015b} 
including ohmic and ambipolar diffusion finding MRI-channel modes to develop
in the upper layers of the disk. They furthermore stress the importance of ambipolar diffusion 
enabling a more laminar wind flow and decreased mass loss rates.
\cite{Bethune2017a} find asymmetric wind flows, nonaccreting configurations
and self-organizing structures such as rings. The results of \cite{suriano_wind_ad_2018} 
corroborate the ones by \cite{Bethune2017a} such that rings emerge in 
ambipolar diffusion dominated disks by reconnection events.
A combination of both photoevaporation and magnetically driven winds,
also called magnetothermal winds
were studied by \cite{wang_2019} and \cite{rodenkirch_2020_wind} 
in the regime of magnetic pressure supported winds. 
With increasing magnetic field strengths, photoevaporative winds 
smoothly transition into magnetic winds driven by the magnetic
pressure gradient in the upper layers of the disk \citep{rodenkirch_2020_wind}.
In contrast to the results of \cite{bai_2017_magnetothermal, suriano_wind_ad_2018, wang_2019} and \cite{rodenkirch_2020_wind},
simulations in \cite{gressel_2020} demonstrate magneto-centrifugal winds mainly
driven by magnetic tension forces. \\
Dust entrainment in protoplanetary disk winds has been mostly examined
in photoevaporative winds. Low gas densities in the wind limit the 
maximum grain size to a few $\si{\micro\m}$. For EUV-photoevaporation
\cite{owen_2011} determined a grain size limit of $\approx \SI{2.2}{\micro\m}$.
\cite{hutchison_2016a} performed small scale smoothed particle hydrodynamics simulations
of a dusty EUV-driven wind and formulated an analytical expression for the 
maximum entrainable dust grain size. They conclude that typically the limiting size
is less than $\SI{4}{\micro\m}$ for typical T Tauri stars and they also argue
that dust settling may lower this limit further.
\cite{franz_2020} simulated dust entrainment in postprocessed XEUV-wind flows
of \cite{Picogna2019} and found a larger maximum dust size of $\SI{11}{\micro\m}$.
Dust entrainment in magnetically driven winds was tested by vertical 1D models 
in the work of \cite{miyake_2016}, concluding that grains in the range of $\SI{25}{\micro\m}$
to $\SI{45}{\micro\m}$ can float up to 4 gas scale heights above the disk mid plane.
\cite{giacalone_2019} used a semianalytical MHD-wind description to 
compute the dust transport and report a maximum dust size of $\SI{1}{\micro\m}$.
Considering the effect of radiation pressure, submicron dust grains
can be blown away from the disk and support clearing out remaining dust \citep{owen_kollmeier_2019}.\\
In this paper we present fully dynamic, multifluid, global, azimuthally symmetric, nonideal MHD simulations 
of protoplanetary disk winds including XEUV-heating to model photoevaporative flows.
We aim to study dust entrainment in magnetically and thermally driven disk winds 
in a dynamic fashion that also considers the vertical dust distribution due to 
turbulent diffusion of dust grains. We thus extend the previous study in \cite{rodenkirch_2020_wind}
by including dust as a pressureless fluid. We furthermore
postprocess the dust density maps to examine observational signatures
of such dusty winds.\\
In Sec.~\ref{sec:model} we describe the theoretical and computational concepts
of the model. The results are presented in Sec.~\ref{sec:results} and 
Sec.~\ref{sec:discussion} discusses limitations of the model as well as 
the comparison to other studies. Finally, in Sec.~\ref{sec:conclusion}
we give concluding remarks.
\begin{table*}[t]
    \label{tab:simulations}
    \caption{Simulations and key parameters}
    \centering
    \begin{tabular}{lccccccccr}
    \hline
    Label & $\beta_0$ & $\alpha$ & X-ray heating & $r_{\mathrm{in}}$ [au] & $r_{\mathrm{out}}$ [au] & $r_\mathrm{c}$ & $\epsilon_\mathrm{ion}$ & Resolution &  Simulation time\\
    \hline
    \hline
    b5c2 &      $10^5$ &   $10^{-4}$ &  \cmark & 1 & 200 & 2    & $10^{3}$ & 500 x 276 &     2000 orbits\\
    b4c2 &      $10^4$ &   $10^{-4}$ &  \cmark & 1 & 200 & 2    & $10^{3}$ & 500 x 276 &     2000 orbits\\
    b4c2np &    $10^4$ &   $10^{-4}$ &  \xmark & 1 & 200 & 2    & $10^{3}$ & 500 x 276 &     1000 orbits\\
    b4c2a3 &    $10^4$ &   $10^{-3}$ &  \cmark & 1 & 200 & 2    & $10^{3}$ & 500 x 276 &     1000 orbits\\
    b4c2eps2 &  $10^4$ &   $10^{-3}$ &  \cmark & 1 & 200 & 2    & $10^{2}$ & 500 x 276 &     1000 orbits\\
    phc2 &      [n/a] & $10^{-4}$ &  \cmark & 1 & 200 & 2    & $10^{3}$ & 500 x 276 &     2000 orbits\\
    phc10 &     [n/a] & $10^{-4}$ &  \cmark & 1 & 200 & 10   & $10^{3}$ & 500 x 276 &     2000 orbits\\
    \hline
    \end{tabular}
    \tablefoot{The numerical resolution is given in number of cells in radial and 
    polar direction. The 
    simulation time is measured in units of orbits at the inner radius $r_\mathrm{in}$.} \label{tab:simulations}
\end{table*}
\section{Model} 
\label{sec:model}
The simulations were carried out with the FARGO3D code \cite{fargo3d,
fargo3d_multi} in a 2.5D axisymmetric setup using a spherical mesh with the
coordinates $(R, \theta, \varphi)$. Throughout this work the cylindrical radius 
will be referred to as $r$, whereas the spherical radius will be denoted 
as $R = \sqrt{r^2 + z^2}$. \revised{
    Observational inclinations, flow angles and viewing angles will 
    be expressed in terms of latitude equivalent to $\theta$ ranging from
    $\SI{90}{\degree}$ to $\SI{-90}{\degree}$.
}
\subsection{Basic equations}
The FARGO3D code solves the following set of equations:
\begin{flalign}
\label{eq:gas_cons_mass}\partial_\mathrm{t} \rho_\mathrm{g} + \nabla \cdot \left( \rho_\mathrm{g} \, \bm{v}_\mathrm{g} \right) &= 0 \,,\\
\label{eq:gas_cons_mom}\rho_\mathrm{g} \left( \partial_\mathrm{t} \bm{v}_\mathrm{g} +  \bm{v}_\mathrm{g} \cdot \nabla \bm{v}_\mathrm{g} \right) &= - \nabla P + \nabla \cdot \bm{\Pi} - \rho_\mathrm{g}\nabla \Phi \\ &+ \frac{1}{4\pi} \left(\nabla \times \mathbf{B}\right) \times \mathbf{B} \,,\\
\label{eq:dust_cons_mass}\partial_\mathrm{t} \rho_\mathrm{d, i} + \nabla \cdot \left( \rho_\mathrm{d, i} \, \bm{v}_\mathrm{d, i} + \bm{j}_\mathrm{i} \right) &= 0  \,,\\
\label{eq:dust_cons_mom}\rho_\mathrm{d, i} \left[ \partial_\mathrm{t} \bm{v}_\mathrm{d, i} +  \left( \bm{v}_\mathrm{d, i} \cdot \nabla \right) \bm{v}_\mathrm{d, i} \right] &= - \rho_\mathrm{d, i}\nabla \Phi + \sum_\mathrm{i} \rho_\mathrm{i} \mathrm{f}_\mathrm{i}  \,,
\end{flalign}
where Eqs.~\ref{eq:gas_cons_mass} and \ref{eq:dust_cons_mass} describe the mass
conservation of gas and dust with $\rho_\mathrm{g}$ and $\rho_\mathrm{d, i}$ being 
the gas and dust density, respectively. The suffix $\mathrm{i}$ denotes the corresponding
dust species. Eqs.~\ref{eq:gas_cons_mom} and \ref{eq:dust_cons_mom} represent 
the momentum equations of gas and dust. We note that we did not consider the dust backreaction
onto the gas in Eq.~\ref{eq:gas_cons_mom} since the dust-to-gas mass ratio is assumed 
to be low. 
The symbol $P$ refers to the thermal gas pressure, $\bm{B}$ the 
magnetic flux density vector, $\bm{v}$ the velocity vectors,
$\Phi = -\frac{GM_*}{R}$ the gravitational potential and $f_\mathrm{i}$ the drag forces
between dust and gas. \\
Turbulent mixing of dust grains caused by the underlying turbulent viscosity in the gas
was considered by defining the diffusion flux $\bm{j}_\mathrm{i}$ \citep{weber_2019}
\begin{equation}
    \bm{j}_\mathrm{i} = -D_\mathrm{i} \left( \rho_\mathrm{g} + \rho_\mathrm{d, i} \right) \nabla \left( \frac{\rho_\mathrm{d, i}}{\rho_\mathrm{g} + \rho_\mathrm{d, i}} \right) \,,
\end{equation}
where the diffusion coefficient $D_\mathrm{i}$ depends on the kinematic viscosity $\nu$ and the Stokes number St \citep{youdin_lithwick_2007}:
\begin{equation}
    D_\mathrm{i} = \nu \frac{1 + \mathrm{St}^2}{\left(1 + \mathrm{St}^2\right)^2} \,.
\end{equation}
The dimensionless Stokes number can be expressed in terms of the stopping time $t_\mathrm{s}$ as follows:
\begin{equation}
\label{eq:stokes_number}  \mathrm{St} = \Omega_\mathrm{K} t_\mathrm{s} = \Omega_\mathrm{K} \frac{\rho_\mathrm{mat} \, a_\mathrm{i}}{\rho_\mathrm{g} \, v_\mathrm{th}} \,,
\end{equation}
with the Keplerian angular frequency $\Omega_\mathrm{K} = \sqrt{G M_* / r^3}$, 
the material density of the dust grain $\rho_\mathrm{math}$ and the mean thermal
velocity of the gas:
\begin{equation}
    v_\mathrm{th} = \sqrt{\frac{8 k_\mathrm{B} T_\mathrm{g}}{\mu m_\mathrm{p} \pi}} \,.
\end{equation}
The Boltzmann constant is expressed as $k_\mathrm{B}$, the gas temperature as $T$, 
the mean molecular weight as $\mu$ and the proton mass as $m_\mathrm{p}$.
The viscosity stress tensor is given by
\begin{equation}
\mathbf{\Pi} = \rho \nu \left[ \nabla \mathbf{v} + (\nabla \mathbf{v})^T - \frac{2}{3} (\nabla \cdot \mathbf{v}) \, \mathbf{I} \right] \,,
\end{equation}
where $\mathbf{I}$ is the unit tensor,
and the energy equation can be stated as
\begin{equation} \label{eq:energy}
\frac{\partial e}{\partial t} + \nabla \cdot (e \mathbf{v}) = - P \, \nabla \cdot \mathbf{v} \,.
\end{equation}
The adiabatic equation of state is given by
\begin{equation}
    P = (\gamma - 1) e \,,
\end{equation}
where $\gamma = 5/3$ is the adiabatic index.
The induction equation including ohmic and ambipolar diffusion reads as follows \citep{Bethune2017a}:
\begin{equation}
\frac{\partial \bm{B}}{\partial t} - \nabla \times \left(\bm{v} \times \bm{B} - 
\frac{4 \pi}{c} \eta_\mathrm{ohm} \cdot \bm{J} + \eta_\mathrm{am} \cdot \bm{J} \times \mathbf{b} \times \mathbf{b} \right) = 0 \,,
\end{equation}
where $\mathbf{b} = \mathbf{B} / |\mathbf{B}|$ the normalized magnetic field vector, $\bm{J}$ the electric current density vector, 
$\eta_\mathrm{ohm}$ the ohmic diffusion coefficient and $\eta_\mathrm{am}$ the ambipolar diffusion coefficient.\\
\begin{figure*}[h]
    \makebox[\textwidth][c] {
      \includegraphics[width=1.0\textwidth]{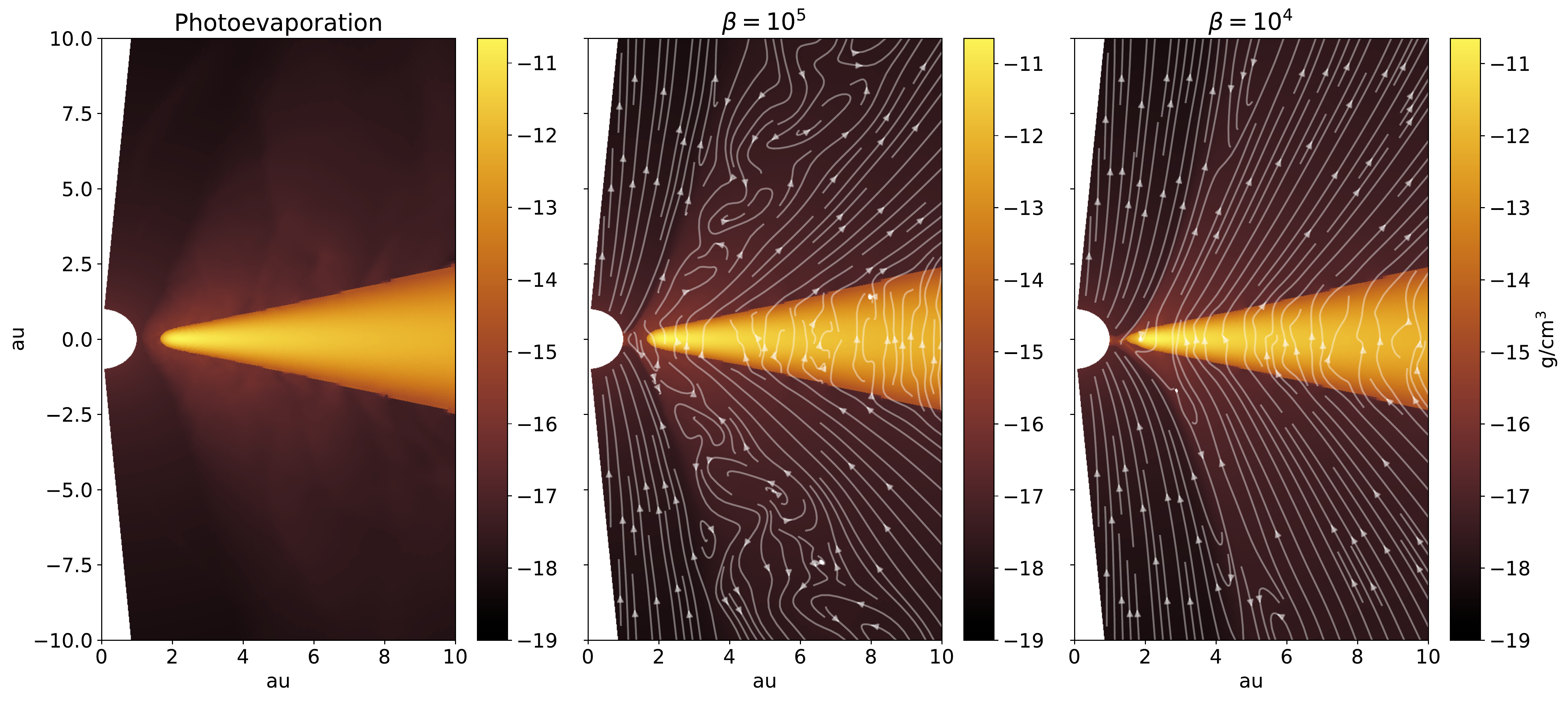}
    }
    \caption[]{Gas density maps of the simulations \texttt{phc2}, \texttt{b5c2} 
    and \texttt{b4c2} after 500 orbits at $\SI{1}{\AU}$. Magnetic field lines are represented by the 
    white lines in the corresponding panels. 
    \revised{The color map scaled logarithmically 
    annotated by the exponent to a base of ten and represents the volume density of the gas
    in $\si{\g\per\cm^3}$.}}
    \label{fig:gas_density_plots}
\end{figure*}
\begin{figure*}[h]
    \makebox[\textwidth][c] {
        \includegraphics[width=1\textwidth]{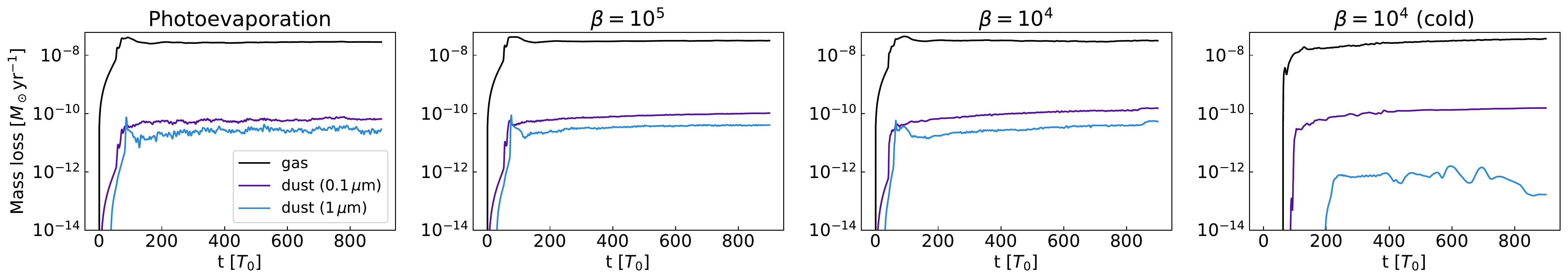}
    }
    \caption[]{Total wind mass loss rates of gas and dust for the simulation
    runs \texttt{phc2}, \texttt{b5c2}, \texttt{b4c2} and \texttt{b4c2np}. Results from 
    $\SI{10}{\micro\m}$ grains are excluded due to the negligible outflow and
    the lack of entrainment in the wind for all runs shown above.}
    \label{fig:mass_losses_time}
\end{figure*}
\begin{figure*}
    \makebox[\textwidth][c] {
        \includegraphics[width=0.85\textwidth]{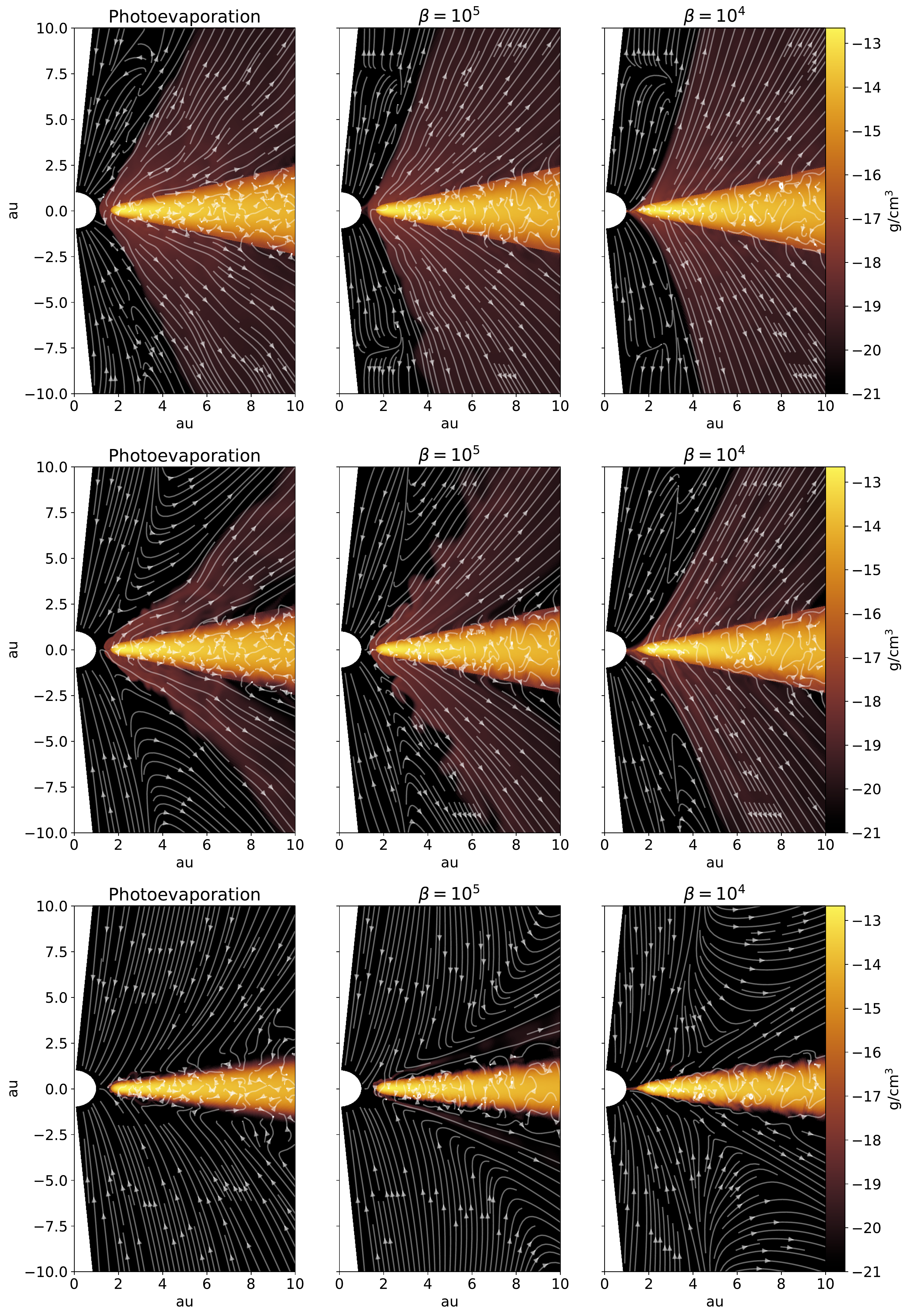}
    }
    \caption[]{Dust density maps for the simulations
    \texttt{phc2}, \texttt{b5c2} 
    and \texttt{b4c2} after 500 orbits at $\SI{1}{\AU}$.
    Dust velocity streamlines are annotated by the white arrowed lines.
    The upper, center and lower three panels visualize the flow of grains 
    with size $\SI{0.1}{\micro\m}$, $\SI{1}{\micro\m}$ and $\SI{10}{\micro\m}$,
    respectively.}
    \label{fig:dust_density_plots}
\end{figure*}
\begin{figure*}
    \makebox[\textwidth][c] {
        \includegraphics[width=0.85\textwidth]{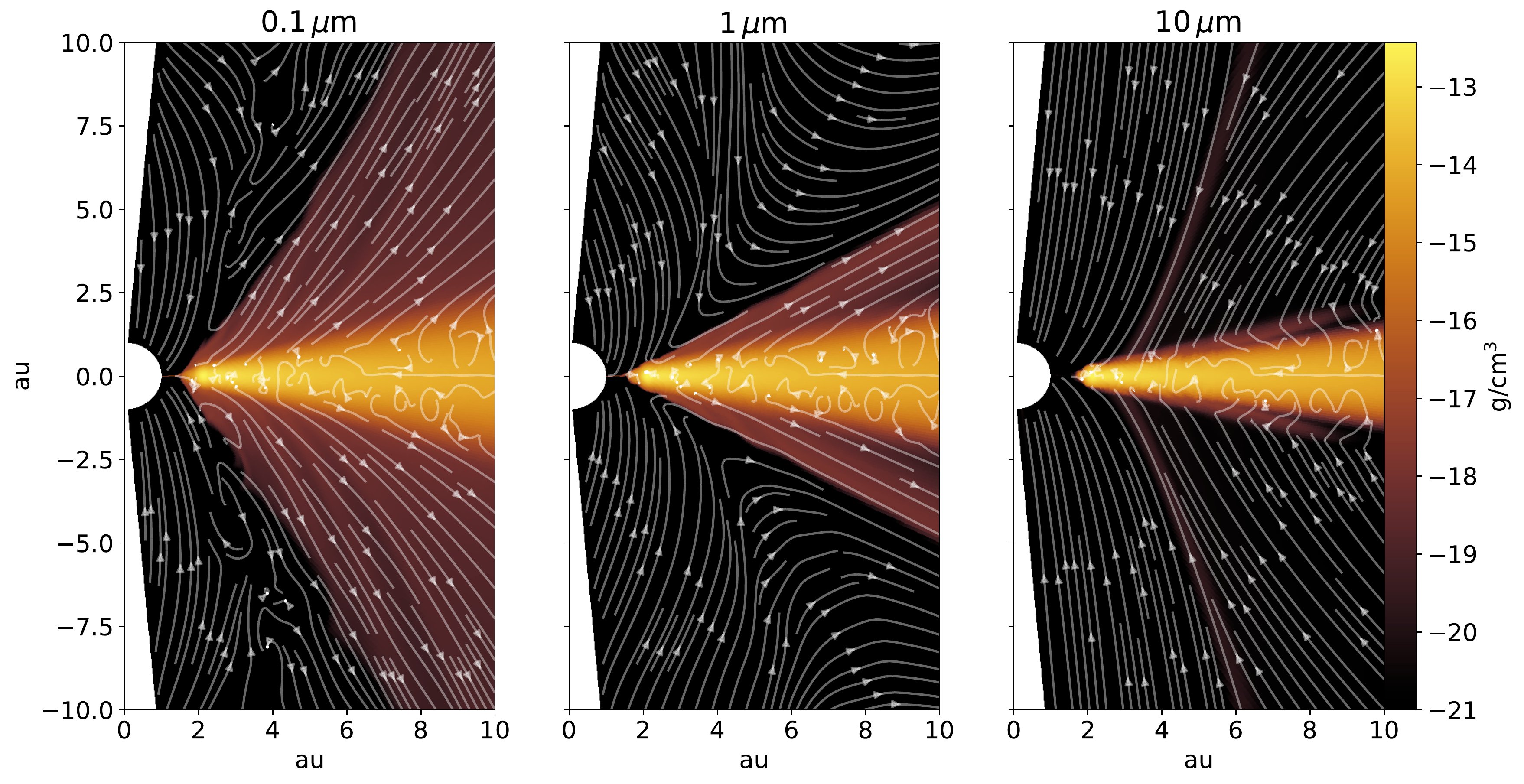}
    }
    \caption[]{Dust density maps for the simulation
    \texttt{b4c2np} after 500 orbits at $\SI{1}{\AU}$.
    Dust velocity streamlines are annotated by the white arrowed lines.
    }
    \label{fig:dust_density_plots_cold}
\end{figure*}
\subsection{Disk model}
\subsubsection{Initial conditions of the gas in the disk}
The vertically integrated surface density profile $\Sigma(r)$ was assumed to be a
power law $\Sigma(r) = \Sigma_0 \, \left(\frac{r}{r_0}\right)^{-\mathrm{p}}$
\revised{with the reference surface density $\Sigma_0$ at the characteristic radius $r_0$
of the system} and the volumetric gas density takes the form
\begin{equation} \label{eq:initial_density}
    \rho(r, z) = \Theta_\mathrm{H}(r - r_\mathrm{c})\frac{\Sigma_0}{\sqrt{2 \pi} H(r)} \left(\frac{r}{r_0}\right)^\mathrm{p} \, \mathrm{exp} \left[\frac{1}{h^2} \left(\frac{r}{\sqrt{r^2 + z^2}} - 1\right)\right] \,,
\end{equation}
where $h = H / r$ is the local aspect ratio and $H = c_\mathrm{s} / \Omega_\mathrm{K}$ the
gas scale height with the local isothermal sound speed $c_\mathrm{s} =
\sqrt{k_\mathrm{B} T(r) / (\mu m_\mathrm{p})}$.
\revised{Within a certain cutoff radius $r_\mathrm{c}$, we set the density to zero with 
a sharp transition and we thus mimiced a disk with an inner cavity of size $r_\mathrm{c}$.
The transition is initially taken care of by the Heaviside function $\Theta_\mathrm{H}(r - r_\mathrm{c})$ in Eq.~\ref{eq:initial_density}.
The sharp transition quickly softens and relaxes to the shape of a smooth inner
rim during the simulation run after a few orbits. The cavity in the model was applied for
both gas and dust.}\\
In a simple blackbody disk being irradiated by stellar light with a grazing
angle $\phi$ onto its surface, the temperature scales with radius as
\citep{chiang_goldreich_1997}
\begin{equation}
    T_\mathrm{g}(r) = \left( \frac{\phi}{2} \right)^{\frac{1}{4}} \left( \frac{R_*}{r} \right)^{\frac{1}{2}} T_* = \left(\frac{\phi L_*}{8 \pi r^2 \sigma_\mathrm{SB}}\right)^{1/4} \,,
    \label{eq:blackbody_disk}
\end{equation}
where $R_*$, $T_*$ and $L_*$ are the stellar radius, temperature and luminosity,
respectively. The Stefan-Boltzmann constant is denoted by
$\sigma_\mathrm{SB}$. Thus, we assumed a radial power law profile for the temperature with $T(r) = T_0 \, r^{-q}$
and the gas scale height scales with radius as $H(r) \propto \sqrt{T(r)} \propto
r^{(3-q)/2}$.\\
The initial values of the azimuthal gas velocity are slightly subkeplerian \citep{nelson_2013}:
\begin{equation}
    v_\phi(r, z) = \Omega_\mathrm{K} r \sqrt{h^2 (p + q)  + (q+1) - \frac{q r}{\sqrt{r^2 + z^2}}} \,.
\end{equation}
We assumed no initial velocities in the radial and polar direction, that is $v_\mathrm{r} = v_\theta = 0$.
\subsubsection{Coronal gas structure} \label{sec:corona}
The hydrostatically stable coronal gas structure with density $\rho_\mathrm{c}(R)$ and pressure $P_\mathrm{c}(R)$ serves as a density floor throughout
the simulation and prevents excessively low initial densities which prevents numerical instabilities.
We used the same prescription as used in the previous models in \cite{rodenkirch_2020_wind} based on \cite{Sheikhnezami2012}.\\
The density was set to
\begin{equation}
\rho_c(R) = \rho_{c0}(r_i) \left( \frac{r_i}{R} \right)^{\frac{1}{\gamma-1}} \,,
\end{equation}
where $r_i$ is the radius at the inner boundary of the simulation domain and 
$\gamma$ the adiabatic index. 
The corresponding pressure profile is given by:
\begin{equation}
P_c(R) = \rho_{c0}(r_i) \frac{\gamma - 1}{\gamma}\frac{GM_\odot}{r_i} 
\left(\frac{r_i}{R} \right)^{\frac{\gamma}{\gamma -1}} \,.
\end{equation} 
The density 
at the inner boundary $\rho_{c0}(r_i)$ is related to the disk's density at the 
midplane and $r_i$ with the density contrast $\delta = \rho_{c0}(r_i) / 
\left[\Sigma(r_\mathrm{i}) / \left(\sqrt{2 \pi} H \right) \right]$. 
Depending on $\Sigma_0$, we set the density contrast to values ranging from $\delta = 
10^{-7}$ to $5 \cdot 10^{-7}$. In absence of X-ray / EUV heating, the gas temperature was directly set to the coronal temperature.
\subsubsection{Dust structure}
Vertical dust settling towards the midplane becomes more effective in the upper layers
of the disk and the overall vertical thickness of the dust layer is over-estimated 
in models of a dust scale height uniquely depending on midplane values as in \cite{dubrulle_dustsettling_1995} \citep{dullemond_dominik_2004_settling}.
We therefore used the following vertical dust distribution
\begin{equation}
    \rho_\mathrm{d}(r, z) = \epsilon_\mathrm{dg} \, \rho(r, z) \left\{\frac{\mathrm{St(r, z=0)}}{\alpha} \left[ \mathrm{exp}\left( \frac{z^2}{2 H^2} \right) - 1 \right] - \frac{z^2}{2 H^2} \right\}
\end{equation}
given in \cite{fromang_2009_dust2}. The dust-to-gas mass ratio of the dust fluids is denoted by $\epsilon_\mathrm{dg}$
and the dimensionless diffusion coefficient follows the $\alpha$-viscosity prescription $\nu = \alpha c_\mathrm{s} H$
\citep{Shakura1973} with the kinematic viscosity $\nu$. We assumed a constant value for the Schmidt number $\mathrm{Sc} = 1$.\\
The initial azimuthal velocities were set to $v_{\mathrm{d}, \phi} =
\Omega_\mathrm{K} \sqrt{r^2 + z^2}$. Similar to the gas, the initial radial and
polar velocities of the dust fluids were set to zero.
\subsubsection{Ionization rate}
In all of the following models X-ray radiation was considered including both a
radial and scattered component. The parametrization was based on the
prescription of \cite{igea_1999_ionization} and \cite{bai_goodman_2009}:
\begin{flalign}
\zeta_{\mathrm{X}}=& \zeta_{\mathrm{X}, \mathrm{sca}}\left[\exp \left(-\frac{\Sigma_{\mathrm{top}}}{\Sigma_{1}}\right)^{0.65}+\exp \left(-\frac{\Sigma_{\mathrm{bot}}}{\Sigma_{1}}\right)^{0.65}\right]\left(\frac{R}{\mathrm{au}}\right)^{-2} \\
&+\zeta_{\mathrm{X}, \mathrm{rad}} \exp \left(\frac{\Sigma_{\mathrm{rad}}}{\Sigma_{2}}\right)^{0.4}\left(\frac{R}{\mathrm{au}}\right)^{-2} \nonumber \,.
\end{flalign}
Both coefficients were multiplied by a factor of 20 compared to the cited values
to model a fiducial X-ray luminosity of $\SI{2e30}{\erg\per\s}$ and were set to
$\zeta_{\mathrm{X}, \mathrm{rad}} = \SI{1.2e-10}{\per\s}$ and 
$\zeta_{\mathrm{X}, \mathrm{sca}} = \SI{2e-14}{\per\s}$. 
The critical column
densities accounted to $\Sigma_1 = \SI{7e23}{\cm^{-2}}$ and $\Sigma_2 =
\SI{1.5e21}{\cm^{-2}}$.\\
\revised{
    The luminosity increase was chosen to be compatible with 
    the photoevaporation model of \citet{ercolano_2008, picogna_2019_photo}
    where the luminosity $L_\mathrm{X} = \SI{2e30}{\erg\per\s}$ was used
    as the fiducial value. Taking into account the mass-luminosity 
    relation presented in \citet{flaischlen_2021}, Fig.~1 based on \citet{getman_2005},
    a luminosity of $L_\mathrm{X} = \SI{1e29}{\erg\per\s}$ would correspond
    in average to a $M_* \approx \SI{0.15}{\Msun}$ star. The increased
    value is thus better suited for the T Tauri star with $M_* = \SI{0.7}{\Msun}$
    assumed in our model. \\
}
Additionally, the cosmic ray ionization rate $\zeta_\mathrm{cr}$ was modeled by the
parametrization of \cite{umebayashi_nakano_2009} where a characteristic
column density of $\SI{96}{\g\,\cm^{-2}}$ was chosen. The symbols 
$\Sigma_\mathrm{bot}$ and $\Sigma_\mathrm{top}$ refer to the column
densities integrated along the polar direction,
starting from the lower boundary (suffix \textit{bot})]
and the upper boundary (suffix \textit{top}).
The radial column densities $\Sigma_\mathrm{rad}$ were computed starting from 
the radial inner boundary.
\begin{flalign}
\zeta_{\mathrm{cr}}=& 5 \times 10^{-18} \mathrm{~s}^{-1} \exp \left(-\frac{\Sigma_{\mathrm{top}}}{\Sigma_{\mathrm{cr}}}\right)\left[1+\left(\frac{\Sigma_{\mathrm{top}}}{\Sigma_{\mathrm{cr}}}\right)^{3 / 4}\right]^{-4 / 3} \\
&+\exp \left(-\frac{\Sigma_{\mathrm{bot}}}{\Sigma_{\mathrm{cr}}}\right)\left[1+\left(\frac{\Sigma_{\mathrm{bot}}}{\Sigma_{\mathrm{cr}}}\right)^{3 / 4}\right]^{-4 / 3} \nonumber
\end{flalign}
To account for short-lived radio nuclides in the disk, an ionization rate of
$\zeta_{\mathrm{nuc}} = \SI{7e-19}{\per\s}$ caused by the $^{26}\mathrm{Al}$
isotope \citep{umebayashi_nakano_2009} acts as an ionization rate floor value.
The resulting total ionization rate then simply sums up to $\zeta =
\zeta_\mathrm{X} + \zeta_\mathrm{cr} + \zeta_\mathrm{nuc}$.
\subsubsection{Ionization fraction} \label{sec:ionization_fraction}
We computed the ionization fraction of the gas via
a semianalytical model for fractal dust aggregates elaborated by
\cite{okuzumi_2009}. The main assumption here is an approximately
Gaussian charge distribution in the dust grains. This approximation is
reasonably well fulfilled if the dust aggregate consists of at least 400
monomers \cite{dzyurkevich_2013}. 
In this approach, the ionization equilibrium was described by the following
equations \cite{okuzumi_2009}:
\begin{equation}
    \frac{1}{1+\Gamma} - \left[ \frac{s_\mathrm{i} u_\mathrm{i}}{s_\mathrm{e} u_\mathrm{e}} \mathrm{exp}(\Gamma)  + \frac{1}{\Theta} \frac{\Gamma g(\Gamma)}{\sqrt{1 + 2 g(\Gamma)} - 1} \right] = 0 \,,
\label{eq:okuzumi_master}
\end{equation}
with the parameter $\Theta$:
\begin{equation}
    \Theta = \frac{\zeta n_\mathrm{g} e^2}{s_\mathrm{i} u_\mathrm{i} \bar{\sigma} \bar{a} n_\mathrm{d}^2 k_\mathrm{B} T} \,.
\end{equation}
The function $g(\Gamma)$ is evaluated as:
\begin{equation}
    g(\Gamma) = \frac{2 \tilde{\alpha} \zeta n_\mathrm{g}}{s_\mathrm{i} u_\mathrm{i} s_\mathrm{e} u_\mathrm{e} (\bar{\sigma} n_\mathrm{d})^2} \frac{\mathrm{exp}(\Gamma)}{1+\Gamma} \,.
\end{equation}
After solving Eq.~\ref{eq:okuzumi_master} numerically for $\Gamma$ with a simple
Newton-Raphson scheme, finally the ion and electron densities $n_\mathrm{i}$ and
$n_\mathrm{e}$ can be calculated:
\begin{equation}
    n_\mathrm{i} = \frac{\zeta n_\mathrm{g}}{s_\mathrm{i} u_\mathrm{i} \bar{\sigma} n_\mathrm{d}} \frac{\sqrt{1 + 2 g(\Gamma)} - 1}{(1 + \Gamma) g(\Gamma)} \,,
\end{equation}
\begin{equation}
    n_\mathrm{e} = \frac{\zeta n_\mathrm{g}}{s_\mathrm{e} u_\mathrm{e} \bar{\sigma} n_\mathrm{d}} \frac{\sqrt{1 + 2 g(\Gamma)} - 1}{\mathrm{exp}(-\Gamma) g(\Gamma)} \,.
\end{equation}
The sticking coefficients $s_\mathrm{i}$
and $s_\mathrm{e}$ were set to 1 and 0.3 respectively, as described in \cite{okuzumi_2009}.
The thermal ion and electron velocities $u_\mathrm{i, e}$ follow:
\begin{equation}
    u_\mathrm{i, e} = \sqrt{\frac{8 k_\mathrm{B} T}{\pi m_\mathrm{i, e}}} \,,
\end{equation}
where $m_\mathrm{e}$ describes the electron mass and 
the ion mass is set to $m_\mathrm{i} = 24 \, m_\mathrm{p}$, corresponding
to magnesium. \\
With the assumption of fractal dust aggregate with a dimension of $D = 2$, the
dust aggregate size becomes $\tilde{a} = a_0 N^{1/2}$ with the number of
monomers $N$ of size $a_0$.  Similarly, the geometric cross section
$\tilde{\sigma}$ of these aggregates can be expressed as $\tilde{\sigma} \approx
\pi a_0^2 N$. The dust number density $n_\mathrm{d}$ can be calculated as
$n_\mathrm{d} = \epsilon_\mathrm{ion} \, n_\mathrm{g} / (\frac{4 \pi}{3} a_0^3 N)$ with the dust-to-gas
ratio $\epsilon_\mathrm{ion}$ used in the ionization model. \\
In the limit of a vanishing dust-to-gas mass ratio the ionization fraction
$x_\mathrm{e}$ approaches
\begin{equation}
    x_\mathrm{e} = \frac{n_\mathrm{e}}{n_\mathrm{g}} = \sqrt{\frac{\zeta}{\tilde{\alpha} n}} \,,
\end{equation}
where $\zeta$ is the total ionization rate, $\tilde{\alpha} = \SI{3e-6}{\cm^{3}\s^{-1}} / \sqrt{T} $ the dissociative
recombination rate coefficient and $n$ the neutral number density of the
molecular species \citep{oppenheimer_dalgarno_1974,ilgner_nelson_2006}.\\
In the same manner as described in \cite{bai_2017_magnetothermal}, we 
increased the ionization fraction in the wind region as a proxy for 
the effect of FUV-radiation:
\begin{equation}
x_\mathrm{fuv} = 2 \cdot 10^{-5} \mathrm{exp}\left[-\left(\frac{\Sigma_\mathrm{rad}}{\Sigma_\mathrm{fuv}}\right)^4 + \frac{0.3 \Sigma_\mathrm{fuv}}{\Sigma_\mathrm{rad} + 0.03 \Sigma_\mathrm{fuv}} \right] \,.
\end{equation}
\subsubsection{Magnetic field}
The magnetic field was set such that the initial plasma beta $\beta$ at the midplane is 
independent of the radius $r$. The poloidal field was initialized from the vector potential $A_\phi$
\begin{equation}
    A_\phi(r, \theta) = \sqrt{\frac{8\pi P}{\beta}} \left(\frac{r}{r_0}\right)^{\frac{2p + q + 1}{4}} \left[1 + m \cdot \mathrm{tan}\left(\theta\right)^{-2}\right]^{-\frac{5}{8}}
\end{equation}
described in \cite{bai_2017_magnetothermal} and exhibits an outward-bent, hourglass shape 
depending on the parameter $m$ where $m = 0$ would correspond to a completely vertical field. 
In order to avoid excessively small time steps in the low density 
regions close to the rotation axis, we limited the local Alfv\'en velocity 
$v_\mathrm{A} = B / \sqrt{4 \pi \rho_\mathrm{g}}$ to $18 v_\mathrm{K}(r_0)$
by increasing the density in the cell accordingly, similar to \cite{Riols2018}.
This procedure does not affect the bulk of the wind flow. 
\subsubsection{Nonideal MHD}
Using the electron fraction $x_\mathrm{e}$, the local ohmic diffusion coefficient
$\eta_\mathrm{ohm}$ can be written as \citep{Blaes1994b}:
\begin{equation}
    \eta_\mathrm{ohm} = \frac{c^2 m_n m_e}{4\pi e^2}\frac{\langle \sigma v \rangle_\mathrm{e}}{x_\mathrm{e}} \approx \frac{234 \, T^{\frac{1}{2}}}{x_\mathrm{e}} \frac{\mathrm{cm}^3}{\mathrm{s}} \,,
\end{equation}
where $m_\mathrm{n}$ and $m_\mathrm{e}$ are the neutral and electron mass
respectively. The electron-neutral collision frequency was set to 
$\langle \sigma v \rangle_\mathrm{e} = 8.28 \cdot 10^{-10} \, T^{-1/2} \mathrm{cm}^3 \mathrm{s}^{-1}$ \citep{draine_1983}.
The ambipolar diffusion coefficient
$\eta_\mathrm{am}$ becomes:
\begin{equation}
    \eta_\mathrm{am} = \frac{B^2}{4 \pi \langle \sigma v \rangle_\mathrm{i} n_\mathrm{g}^2 x_\mathrm{i}} \,.
\end{equation}
The ion-neutral collision rate was set to
$\langle \sigma v \rangle_\mathrm{i} = 1.9 \cdot 10^{-9} \mathrm{cm}^3 \mathrm{s}^{-1}$ \citep{draine_1983}.
Since large diffusivities severely limited the simulation time step 
a super time stepping scheme was implemented and more details are 
given in Appendix~\ref{sec:sts_appendix}.
We furthermore limited the magnetic Reynolds number for ambipolar diffusion 
$Rm = c_\mathrm{s} \, H / \eta_\mathrm{am}$ to $Rm \ge 0.05$, similar 
to \cite{Gressel2015b}, confirming that this approximation does not 
significantly alter the simulation outcome.
\subsubsection{Heating and cooling}
We employed the same temperature presciption as described in 
\cite{rodenkirch_2020_wind} based on \cite{picogna_2019_photo} to mimic the
outflow due to photoevaporation in the simulation runs modeling a warm
(magneto-) photoevaporative wind. According to their model, XEUV heating
is
significant up to radial column densities of $\Sigma_\mathrm{r, crit} = \SI{2.5e22}{\cm^{-2}}$. In these
regions the temperatures $T_\mathrm{photo}$ were updated to the values given by Eq.~1 in \cite{picogna_2019_photo}.
Beyond the critical column density $\Sigma_\mathrm{r, crit}$ the temperature was cooled to
the original gas temperature $T_0 = T_\mathrm{g}$. \\
In the case of the cold magnetic wind model \texttt{b4c2np} (see Tab.~\ref{tab:simulations}), the temperature in the corona 
was only slightly increased by 20 \% compared to the bulk temperature 
in the disk.
We used a simple $\beta$-cooling recipe in the whole simulation domain to
exponentially damp the temperatures to $T_0$ on a time scale $\beta_\mathrm{cool}$
\begin{equation}
  T(t^\mathrm{n} + \Delta t) = T(t^\mathrm{n}) + \left( T(t^\mathrm{n}) - T_0 \right) \mathrm{exp} \left(\frac{\Delta t}{\beta_\mathrm{cool}} \right) \,.
\end{equation}
\subsection{Boundary conditions}
We employed simple symmetric boundary conditions for most of the variables and
boundaries where the values of the active cells are copied to the ghost cells.
At the inner radial boundary, the radial velocity followed an outflow boundary condition
where $v_\mathrm{r} = 0$ in the ghost zone if $v_\mathrm{r} > 0$ in the active cell.
In the contrary case, $v_\mathrm{r}$ was copied to the ghost cell. The same procedure 
was applied to the outer radial boundary with the opposite sign.
With this formulation mass influx from outside of the domain was avoided.\\
At the $\theta_\mathrm{min}$ and $\theta_\mathrm{max}$ boundaries $v_\phi$, 
$v_\theta$, $B_\phi$, $B_\theta$ and the EMF in radial direction change sign in
the ghost zone to mimic a polar boundary.
In MHD disk models the radial inner boundary condition can lead to spurious effects
in the simulation domain. The same applies to the outer radial boundary where, if ill
defined, artificial collimation affects the flow \citep{ustyugova_1999_boundary}.
To avoid these difficulties we added a zone with additional artificial ohmic diffusion,
so that the coupling between the gas and the magnetic field is weak enough that 
spurious effects were damped and the simulation domain remained largely unaffected. 
A similar approach was used at the inner boundary in \cite{cui_2021_ambipolar_3d} 
to stabilize the inner region of the domain. \\
Dimensionless values of the artificial ohmic resistivity ware $\eta_\mathrm{ohm,
in} = \num{2e-4}$ and $\eta_\mathrm{ohm, out} = \num{1e1}$ at the inner and
outer damping zone, respectively.  The extent of the damping zones is 10\% of
the inner and outer radius. With the choice of these values we found that the
simulation domain and the timestep are largely unaffected by the boundary
conditions. Adding an inner cavity with greatly reduced amount of gas and dust
furthermore ensures negligible impact of the inner boundary condition on the 
wind flow.
\subsection{Radiative transfer model and post-processing} \label{sec:radtrans}
We post-processed the dust density outputs with the radiative transfer code
RADMC-3D \citep{dullemond_radmc3d_2012}. 
In order to obtain three dimensional dust density distributions
the axisymmetric simulation outputs were extended in the azimuthal direction
by repeating the densities in 300 cells in $\phi$.
For the thermal Monte-Carlo step and the image reconstruction with scattered light
$n_\mathrm{phot} = 10^6$ and $n_\mathrm{phot\_scat} = 10^8$ photon packages
were used, respectively.\\
\revised{
The thermal Monte-Carlo calculation was based on a modified version of the 
Bjorkman-Wood method \citep{bjorkman_wood_2001} and the resulting 
temperatures were used in the synthetic images instead of the 
temperature profile prescribed in the MHD simulations.
Every synthetic observation was created with only one dust species
since the limited number of fluid would lead to artificial discrete step-like 
features in the image. Only considering one dust species at a time
may lead to unrealistic temperature profiles but after checking the 
output of the thermal calculation of RADMC-3D the temperatures 
agreed with the dust temperature estimated computed in Appendix~\ref{sec:dust_appendix}
in the relevant regions. We further emphasize that the synthetic 
images were produced in the H-band at a wavelength of $\SI{1.6}{\micro\m}$
where scattered light effects dominate and thermal emission from
the dust grains at these low temperatures is negligible.
}
\\
The dust opacities were computed with \texttt{optool} \citep{dominik_optool}
and are similar to the DIANA opacities \citep{woitke_diana_2016}
with a material mixture of 87 \% pyroxene, 13 \% carbon (by mass)
but with an increased porosity of 64 \% to be consistent with 
a dust material density of $\rho_\mathrm{mat} = \SI{1}{\g\per\cm^3}$ in
the simulations. Concerning the dust grain sizes, we computed the opacities for  $a_1 = \SI{0.1}{\micro\m}$, $a_2 = \SI{1}{\micro\m}$ 
and $a_3 = \SI{10}{\micro\m}$ corresponding to the dust sizes in the simulations.
Anisotropic scattering effects were included by using the 
Henyey-Greenstein phase function approximation \citep{henyey_greenstein_1941}.
A classical T Tauri star (CTT) was assumed with a temperature of $T_* = \SI{4000}{\kelvin}$ and
a radial extent of $R_* = \SI{2.55}{\Rsun}$
at a distance of $\SI{150}{\parsec}$ which is in the 
range of typical properties of observed CTTs \citep{alcala_2021}.
The resulting images were convolved with a gaussian beam size of 75 mas
to account for typical observational resolution limits using VLT / SPHERE
\citep{sphere_2019,avenhaus_2018}.
\begin{figure*}[h]
    \makebox[\textwidth][c] {
        \includegraphics[width=1\textwidth]{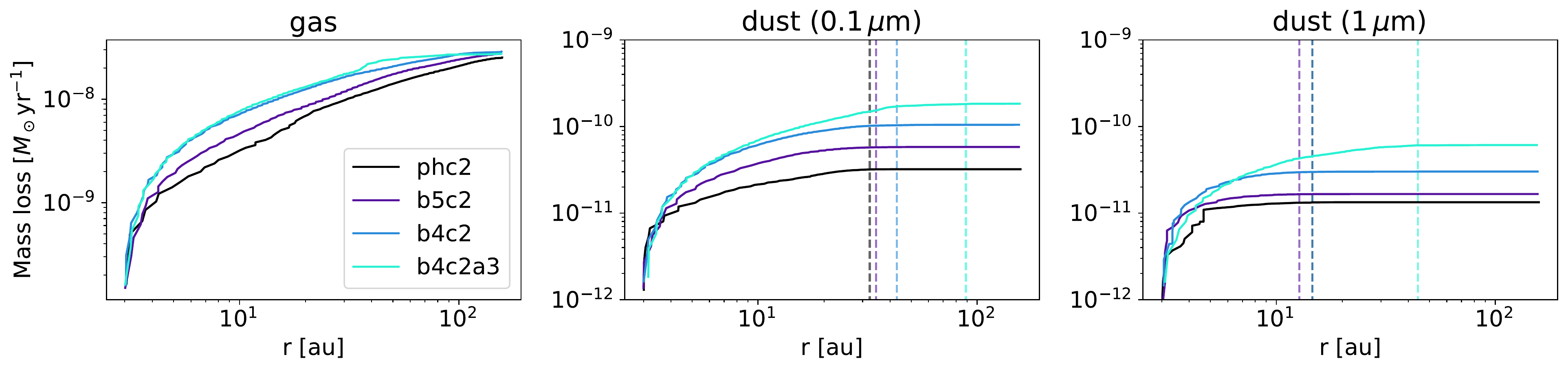}
    }
    \caption[]{Cumulative mass loss rates of gas and dust for the simulation
    runs \texttt{phc2}, \texttt{b5c2}, \texttt{b4c2} and \texttt{b4c2a3} starting from 3~au.
    The cylindrical radius or foot point of the loss is obtained by tracing
    backwards the corresponding streamline from the outer radial boundary down
    to 5 scale heights above the disk mid plane. The flow field is averaged over
    50 orbits at 1 au starting at 600 orbits. The vertical dashed lines mark the limit 
    of the region where 99 \% of the mass loss occurred.}
    \label{fig:massloss_radial}
\end{figure*}

\section{Results} \label{sec:results}
Sec.~\ref{sec:parameters} describes the relevant parameters and simulation runs.
In the following subsections we present the simulation results starting with a
description of the wind flow structure emerging from the disk in
Sec.~\ref{sec:gas_flow_structure} and continue with the dust dynamics 
in Sec.~\ref{sec:dust_flow_structure}. 
The effectiveness of dust entrainment and
the outflow angle of the grains is addressed in
Sec.~\ref{sec:flow_angle}, Sec.~\ref{sec:launching_surface} and Sec.~\ref{sec:streamlines}.  
Finally, in Sec.~\ref{sec:synthetic_observations} we compare synthetic
observations computed from the simulation output with existing observations. 
\subsection{Parameters \& normalization} \label{sec:parameters}
Tab.~\ref{tab:simulations} summarizes the key parameters
of the different simulation runs presented in this study.
The first part of the label describes wether a magnetic wind (for instance \texttt{b4} with $\beta = 10^4$) 
or a photoevaporative wind without magnetic fields 
was modeled (with label \texttt{ph}). Subsequently, the letter \enquote{c} refers to the cavity location 
(such as \texttt{c2} for a cavity at $\SI{2}{\AU}$). The sublabels \texttt{np}, \texttt{a3} and \texttt{eps2}
refer to \enquote{no photoevaporation}, $\alpha = 10^{-3}$ and $\epsilon_\mathrm{ion} = 10^{-2}$, respectively.\\
In the code all quantities were treated in dimensionless units that 
are scaled as follows:
\begin{flalign}
    v_0 &= r_0 \, \Omega_K(r_0)\,,\\
    \rho_0 &=  M_\odot / r_0^3\,,\\
    P_0 &= \rho_0 v_0^2 \,,
\end{flalign}
where the length unit was set to $r_0 = \SI{5.2}{\AU}$, \revised{the semi-major axis
of Jupiter as a natural scale of stellar systems}.
The initial density profile was scaled to reach a surface density of 
$\Sigma = \SI{200}{\g\per\cm^2}$ at $r = \SI{1}{\AU}$ and decreases radially
with power law slope of $p = 1$. 
The central star mass was set to $\SI{0.7}{\Msun}$ and the disk 
follows a moderate aspect ratio of $H / r_0 = 0.055 \, (r / r_0)^{1/4}$ 
which corresponds to a flared disk with a temperature slope of $q = 1/2$. 
These parameters correspond to a stellar luminosity of $L_* = \SI{1.5}{\Lsun}$
with the assumption of a grazing angle of $\phi = 0.02$.
The luminosity is in line with the stellar parameters of $T_* = \SI{4000}{\kelvin}$
and $R_* = \SI{2.55}{\Rsun}$ mentioned in Sec.~\ref{sec:radtrans}.
\\
A temperature floor of $T_\mathrm{min} = \SI{10}{\kelvin}$ and a cooling 
time scale of $\beta_\mathrm{cool} = 10^{-4}$ was applied.
The fiducial value of the inner cavity size was chosen to be $r_\mathrm{c} = \SI{2}{\AU}$.
\revised{Further parameters are the adiabatic index of $\gamma = 5 / 3$
corresponding to an ideal atomic gas and 
the mean molecular weight $\mu = 1.37125$, consistent with 
the value chosen in \citet{picogna_2019_photo} for the composition of the ISM \citep{hildebrand_1983}. \\}
In Sec.~\ref{sec:ionization_fraction} the model necessitates 
an assumption of the dust content in the disk. In this work 
the ionization model was parametrized with a static value of the dust-to-gas
ratio $\epsilon_\mathrm{ion} = 10^{-3}$ and a monomer size of $a_0 = \SI{5}{\micro\m}$
that constitute aggregates with 400 particles, resulting in a grain size of 
$\SI{100}{\micro\m}$. Increasing the dust-to-gas ratio by one order of magnitude
did not alter the ionization fraction profile by a great amount, as 
described in Appendix~\ref{sec:ion_appendix}. In order to simplify 
the model, the dynamically simulated dust fluids were not coupled 
to the ionization model and the relative insensitivity to the chosen parameters
in the midplane supports this approximation. \\
All simulation runs were carried out with three dust species 
of the sized $a=\SI{0.1}{\micro\m}$, $\SI{1}{\micro\m}$ and $\SI{10}{\micro\m}$,
intended to represent the three cases of full entrainment, slow decoupling, and
fast decoupling from the gas in the wind.
The material density of the grains was set to $\rho_\mathrm{mat} = \SI{1}{\g\per\cm^3}$.\\
The computational grid adopts a spherical symmetry and contains 500 logarithmically spaced cells in radial
direction whereas 276 cells were used in the polar ($\theta$) direction.
In order to increase the vertical resolution closer to the disk mid plane, the
polar cells were arranged in a stretched grid, with a cell size at the polar boundaries
that was twice as larger compared to the mid plane.
With this approach a resolution of 7 cells per $H$ in polar direction at $r = r_0, z = 0$
was reached.
\revised{
To avoid numerical issues at the polar boundaries the grid limits in this direction
were chosen such that $\theta_\mathrm{min} \approx \SI{87}{\degree}$
and $\theta_\mathrm{max} \approx \SI{-87}{\degree}$. 
The radial domain spans from $\SI{1}{\AU}$ to $\SI{200}{\AU}$ as
indicated in Tab.~\ref{tab:simulations}.
}

\subsection{Gas flow structure} \label{sec:gas_flow_structure}
Fig.~\ref{fig:gas_density_plots} visualizes the gas wind flow for the simulation runs
\texttt{phc2}, \texttt{b5c2} and \texttt{b4c2}. The latter two include magnetic fields
whereas the former only includes the photoevaporation recipe.
Comparing both magnetic winds in Fig.~\ref{fig:gas_density_plots} the 
magnetic field structure consists mostly of straight field lines for $\beta = 10^4$, 
whereas the weaker field shown in the central panel exposes a more irregular field structure
towards the inner region of the disk. For this field strength and configuration
the wind is in the transition between a thermally dominated and a magnetically dominated
wind. We note that these are simulation snapshots and not averaged wind flows. \\
In terms of wind total mass loss rates all of these three runs display similar 
results, as shown in Fig.~\ref{fig:mass_losses_time}. Generally, 
the wind converges to a steady-state flow after a time scale about 200 orbits at $\SI{1}{\AU}$
and the simulation time chosen for these runs is thus sufficient to explore relevant
wind dynamics in the inner region. \\
The sharp cutoff of the disk at $r = \SI{2}{\AU}$ quickly relaxes towards
a stable equilibrium after a few orbits. The choice of the initial shape of the inner rim 
is thus not relevant for the wind flow.
We observe accretion of gas through the inner cavity when a magnetic field is present.
This effect, which can be best observed in the right panel of Fig.~\ref{fig:gas_density_plots},
is more pronounced for larger magnetic field strengths and is caused
by wind-driven accretion.\\ 
The photoevaporation gas mass loss rates reach $\approx \SI{2.83e-8}{\Msun\per\year}$,
in good agreement with \cite{picogna_2019_photo}. The mass loss rates in the simulation runs
\texttt{b5c2} and \texttt{b4c2} are 
$\approx \SI{3.13e-8}{\Msun\per\year}$ and $\approx \SI{3.03e-8}{\Msun\per\year}$,
respectively. In the case of a purely magnetically driven wind without any heating
by ionizing radiation of model \texttt{b4c2np} the wind mass loss rates are similar
with a value of $\SI{3.19e-8}{\Msun\per\year}$.
The mass loss rates were computed at 95 \% of the outer radial simulation boundary.
The region within five gas scale heights of the disk was excluded from the mass loss rate
to avoid the corruption of the results from circulation within the disk.
\subsection{Dust flow structure} \label{sec:dust_flow_structure}
In Fig.~\ref{fig:dust_density_plots} the dust flows are shown for the grain sizes
$\SI{0.1}{\micro\m}$,
$\SI{1}{\micro\m}$ and
$\SI{10}{\micro\m}$. Generally, $\SI{10}{\micro\m}$ sized grains are hardly
entrained with the wind flow since the drag force is too weak to counter act the
gravitational pull. $\SI{0.1}{\micro\m}$ grains are well coupled to the gas
flow and can be considered as tracers of the wind. Throughout the analysis we
refer to the flow angle or inclination angle, meaning the angle enclosed
between the dust flow direction and this disk mid plane axis at $z = 0$. 
\revised{The
inclinations $\theta_\mathrm{d}$ of the dust flows are smaller in terms of latitude for the thermally
driven wind shown on the left-hand side compared to the magnetically driven wind
on the right-hand side. The magnetic field is well coupled to the gas in the
wind region where the ionization fraction $x_\mathrm{e, i}$ surpasses $10^{-4}$
and the field topology forces the wind flow to be more inclined compared to the
thermally driven wind.}\\
The same phenomenon also appears for $\SI{1}{\micro\m}$ grains shown in the
second row of Fig.~\ref{fig:dust_density_plots}. Due to the weaker coupling and
the lack of pressure support the dust flow is less inclined with respect to the
gaseous wind compared to smaller grains. The dust flow exhibits wave-like
patterns in the case of $\beta = 10^5$ and becomes more stable for larger
magnetic field strengths. These wave structures coincide with the irregular
magnetic field patterns displayed in Fig.~\ref{fig:gas_density_plots}. We
furthermore point out that the dust flow consisting of $\SI{1}{\micro\m}$ grains
is efficiently launched in the inner part of the disk, whereas beyond several au
the wind significantly weakens. The dusty wind cannot be sustained in the lower
density regions further out and up in the disk since the Stokes number
significantly increases.\\
An example of a dust flow in a cold magnetically driven wind is given
in Fig.~\ref{fig:dust_density_plots_cold} for an initial $\beta = 10^4$.
The inclination with respect to the mid plane is smaller compared to 
the warm magnetothermal or photoevaporative winds.
Grains with a size of $\SI{1}{\micro\m}$ are entrained at a rather shallow
inclination angle and similar to the warm winds, dust entrainment is negligible
considering $\SI{10}{\micro\m}$ dust particles.
While material in the winds including photoevaporation is ejected with 
a velocity of $\approx \SI{15}{\km\per\second}$ the outflow velocity of the cold magnetic wind
in \texttt{b4c2np} only reaches values of $\approx \SI{5}{\km\per\second}$.
The decreased wind speed causes a smaller drag force between gas
and thus lowers the dust entrainment efficiency.
Given that the mass loss rate is approximately equal to the one of the 
heated winds, the gas and dust densities in the wind region are significantly higher.\\
In Fig.~\ref{fig:massloss_radial} the cumulative mass loss rate depending on the cylindrical
radius of the foot point of the wind is shown. The wind streamlines were traced backwards
from the cells close to the outer radial boundary and the foot point was registered when 
the streamlines crosses the surface at 5 gas pressure scale heights.
Only streamlines with foot point $r_\mathrm{f} \ge \SI{3}{\AU}$ were considered since
the wind flow structure and the limited resolution did not in every case ensure a
successful construction of traceable streamlines close to the inner cavity.
The mass loss rates were obtained from a velocity and density field averaged over
50 orbits at 1 au starting from 600 orbits. \\
Warm magnetic winds lead to more dust entrainment
and a stronger magnetic field increases the dust mass loss rate significantly.
The dashed lines in Fig.~\ref{fig:massloss_radial} mark the radius within 
from within 99 \% of the mass loss occurs, which illustrate less efficient
dust entrainment for larger grains further out in the disk. Numerical values
are $\SI{32.3}{\AU}$, $\SI{34.5}{\AU}$, $\SI{43}{\AU}$ and $\SI{88.8}{\AU}$ for the simulations \texttt{phc2}, \texttt{b5c2},
\texttt{b4c2}, \texttt{b4c2a3} and $\SI{0.1}{\micro\m}$ grains, respectively. Analogously, 
the limits for $\SI{1}{\micro\m}$ grains are $\SI{14.6}{\AU}$, $\SI{12.7}{\AU}$, $\SI{14.6}{\AU}$ and $\SI{44.3}{\AU}$.
The general trend is that the dust grains are preferably entrained starting 
from the inner regions of the disk. The limiting radius decreases with increasing
dust size and hence we expect a much larger ejection region of dust for the smallest grain size. 
Considering the results of the simulation \texttt{b4c2a3} with
an increased viscosity of one order of magnitude ($\alpha = 10^{-3}$)
the limiting radius lies significantly further outward. 
Increased turbulent diffusion effectively transports dust grains towards
the wind launching front at larger radii.\\
Going back to Fig.~$\ref{fig:mass_losses_time}$ the mass loss rate for the two
relevant dust species stabilizes over a few hundred orbits in the case of a
thermal wind. In the case of magnetically driven winds however, dust mass loss
rates reach a quasi-steady state but are slowly increasing throughout the
simulation. In the presence of a cold magnetic wind, the mass loss rates of
larger micron-sized dust grains are lower compared to the ionized winds since
the flow angle is rather shallow and the entrainment efficiency suffers from the
decreased wind speed.
\\
\begin{figure*}[h]
    \makebox[\textwidth][c] {
        \includegraphics[width=1.0\textwidth]{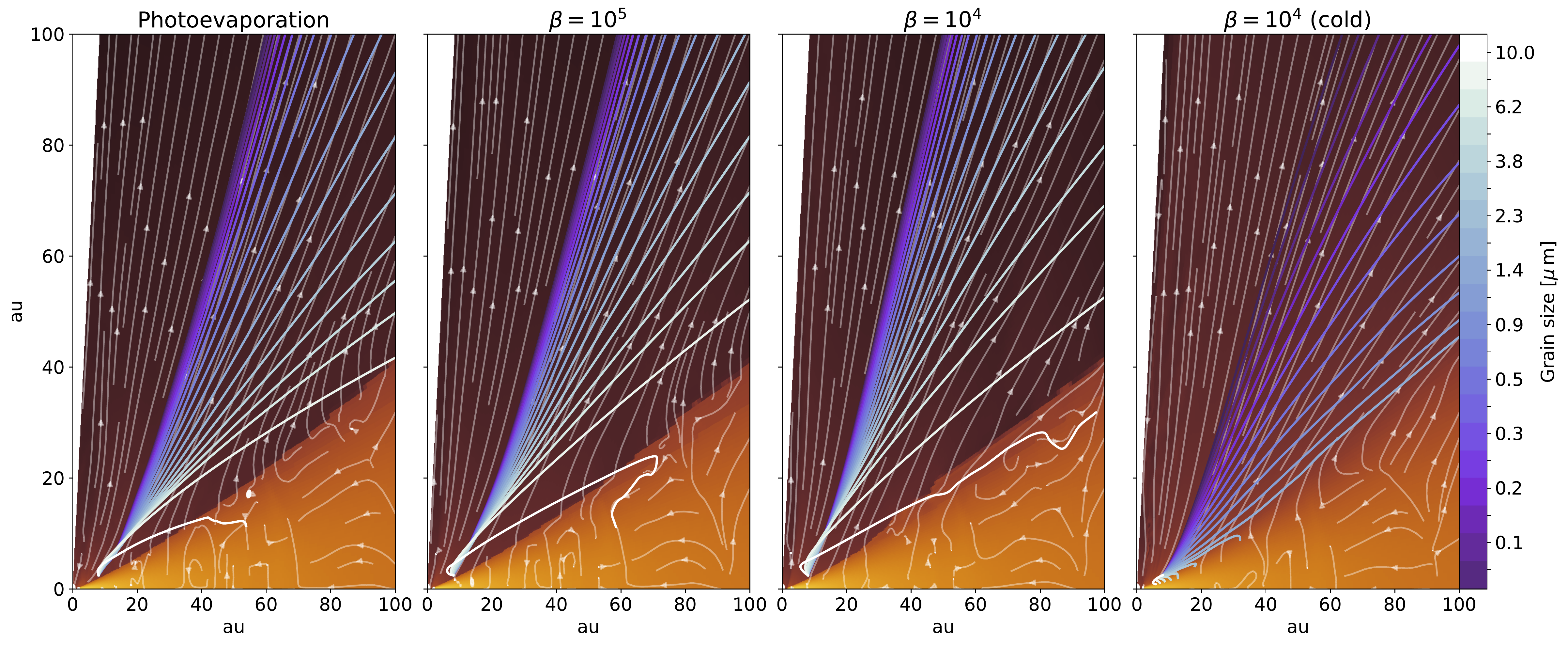}
    }
    \caption[]{Numerically integrated dust trajectories for snapshots after 500 $T_0$
    of the simulation runs \texttt{phc2}, \texttt{b5c2} 
    and \texttt{b4c2}. The orange color map represents the gas density and the thin arrowed white
    lines denote the gas velocity streamlines. The thicker colored lines visualize 
    the dust trajectories for grain sizes ranging from $\SI{0.1}{\micro\m}$ to 
    $\SI{10}{\micro\m}$.}
    \label{fig:dust_trajectories_semi}
\end{figure*}
\subsection{Maximum grain size and flow angle} \label{sec:flow_angle}
In order to quantify the maximum entrained grain size and the entrainment angle, 
we chose the approach to numerically integrate streamlines of various dust species
based on the gas velocity field from the simulations. 
The following system of equation was solved numerically with the given
velocity field of the corresponding simulation output:
\begin{flalign}
    v_\mathrm{d, r} &= v_{\mathrm{d}, \phi} - \frac{GM_*}{\left(r^2 + z^2\right)^\frac{3}{2}} r - \frac{v_\mathrm{d, r} - v_\mathrm{g, r}}{t_\mathrm{s}}\,, \\
    v_\mathrm{d, z} &= - \frac{GM_*}{\left(r^2 + z^2\right)^\frac{3}{2}} z - \frac{v_\mathrm{d, z} - v_\mathrm{g, z}}{t_\mathrm{s}} \,.
\end{flalign}
The numerical integration was carried out with the VODE solver \cite{brown_vode_1989}.
We verified that the streamlines were comparable to the ones in the dust velocity 
field output from the simulation. No significant deviations were visible.\\
The results are shown in Fig.~\ref{fig:dust_trajectories_semi} for 20
logarithmically distributed grain sizes ranging from $\SI{0.1}{\micro\m}$
to $\SI{10}{\micro\m}$. The resulting trajectories match the corresponding 
dust streamlines from the actual simulation.
As the starting point we chose $r = \SI{8}{\AU}, z = \SI{2.5}{\AU}$ which is
situated well above the wind launching front. As expected from the simulation
outputs $\SI{10}{\micro\m}$ grains fall back towards the disk surface in the
case of winds including photoevaporation (models \texttt{phc2}, \texttt{b5c2}
and \texttt{b4c2}). Generally, grains are lifted more efficiently in the case
of warm, magnetic winds. Grains larger than $\approx \SI{3}{\micro\m}$ to
$\SI{4}{\micro\m}$ are lifted further but eventually fall back onto the disk at
larger radii in the presence of a thermal wind. \\ For warm, magnetically driven
winds, the limiting grain size shifts towards $\approx \SI{6}{\micro\m}$. In a
cold, magnetic wind the picture changes and only submicron dust grains are
lifted far from the disk surface and particles larger than $\SI{1}{\micro\m}$
fall back onto the disk or are not entrained at all. The decreased wind speed
is responsible for the lower dust entrainment efficiency as mentioned in
Sec.~\ref{sec:dust_flow_structure}.
\\
\begin{figure}
    \centering
    \includegraphics[width=0.5\textwidth]{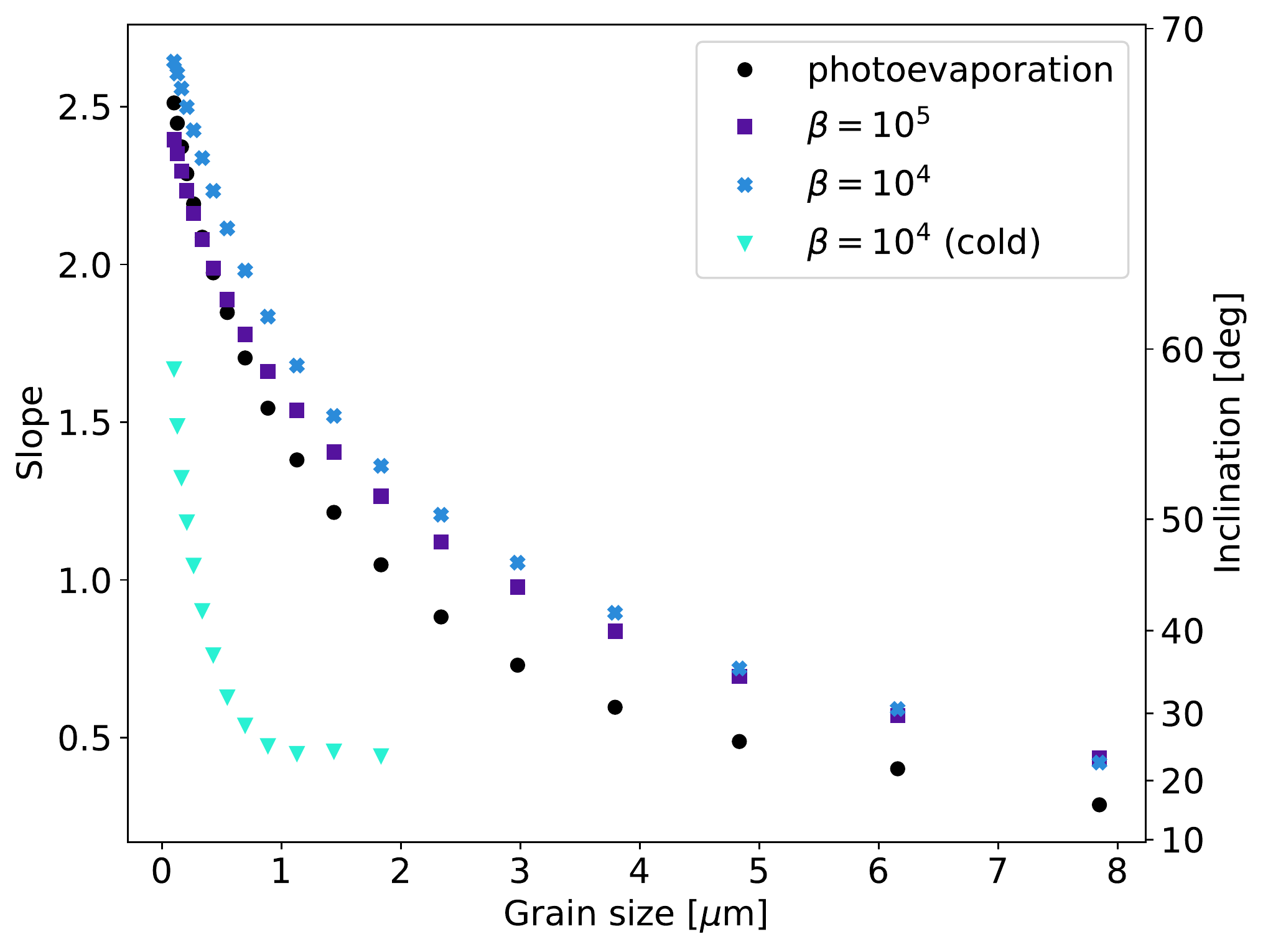}
    \caption[]{Asymptotic dust flow inclination depending on the grain size 
    for the simulation runs \texttt{phc2}, \texttt{b5c2}, \texttt{b4c2} and \texttt{b4c2np}.}
    \label{fig:dust_inclination}
\end{figure}
Since the dust flow inclination angle is strongly dependent on the grain size,
we quantified the inclination by fitting linear functions to the dust trajectories
presented in Fig.~\ref{fig:dust_trajectories_semi}. We only considered the final
25 \% part of the cylindrical radius of the trajectory for the fits. 
The corresponding slopes and inclination angles based on the simulations
\texttt{phc2}, \texttt{b5c2}, \texttt{b4c2} and \texttt{b4c2np} are plotted in
Fig.~\ref{fig:dust_inclination}.
For small grains and warm winds the difference in the inclination angle is
rather small.
The inclination of the flow of $\SI{1}{\micro\m}$ grains ranges from
roughly $\SI{67}{\degree}$ to $\SI{69}{\degree}$. Considering $\SI{3}{\micro\m}$
grains however, the inclination angle is $\SI{36.1}{\degree}$ assuming a 
thermal wind compared magnetic winds with $\SI{44.4}{\degree}$ and $\SI{46.5}{\degree}$
for $\beta = 10^5$ and $\beta = 10^4$, respectively.
Generally, the inclination angle is larger for warm, magnetic winds, especially
considering micron sized grains.\\
A significant difference in inclination is however observed in a cold 
magnetic wind. While $\SI{0.1}{\micro\m}$-sized dust grains 
are entrained by an angle of $\SI{59.1}{\degree}$, the inclination drops
down rapidly to roughly $\SI{25}{\degree}$ for micron-sized grains, as expected.
\begin{figure}
    \centering
    \includegraphics[width=0.5\textwidth]{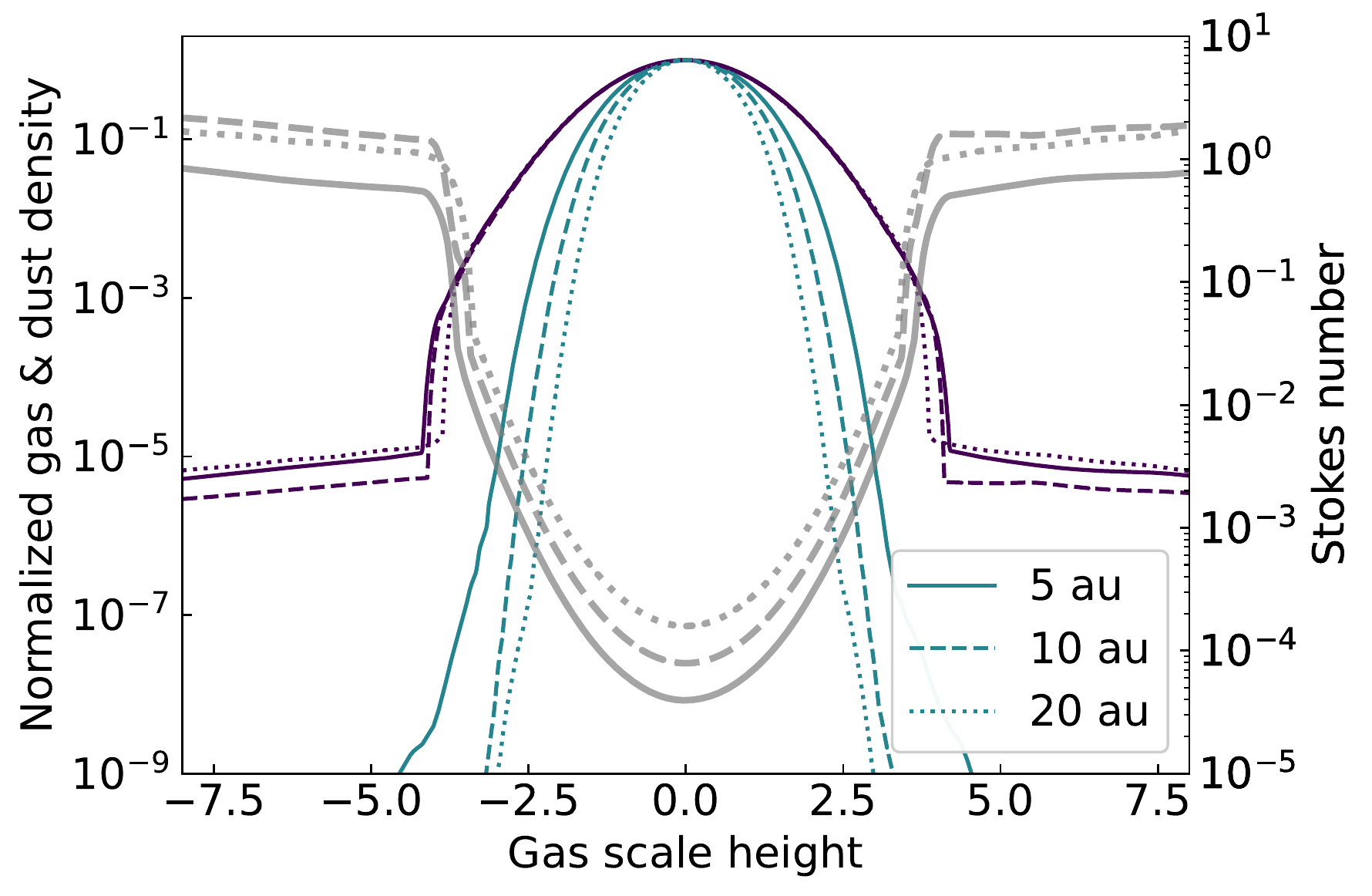}
    \caption[]{Normalized vertical density slices of the \texttt{phc2}
    at different radial locations in the disk. The violet lines represent the normalized
    gas density while the turquoise lines display the dust density of $\SI{10}{\micro\m}$
    grains. The gray lines denote the respective Stokes number profile.}
    \label{fig:slices_radius}
\end{figure}
\begin{figure}[h]
    \centering
    \includegraphics[width=0.5\textwidth]{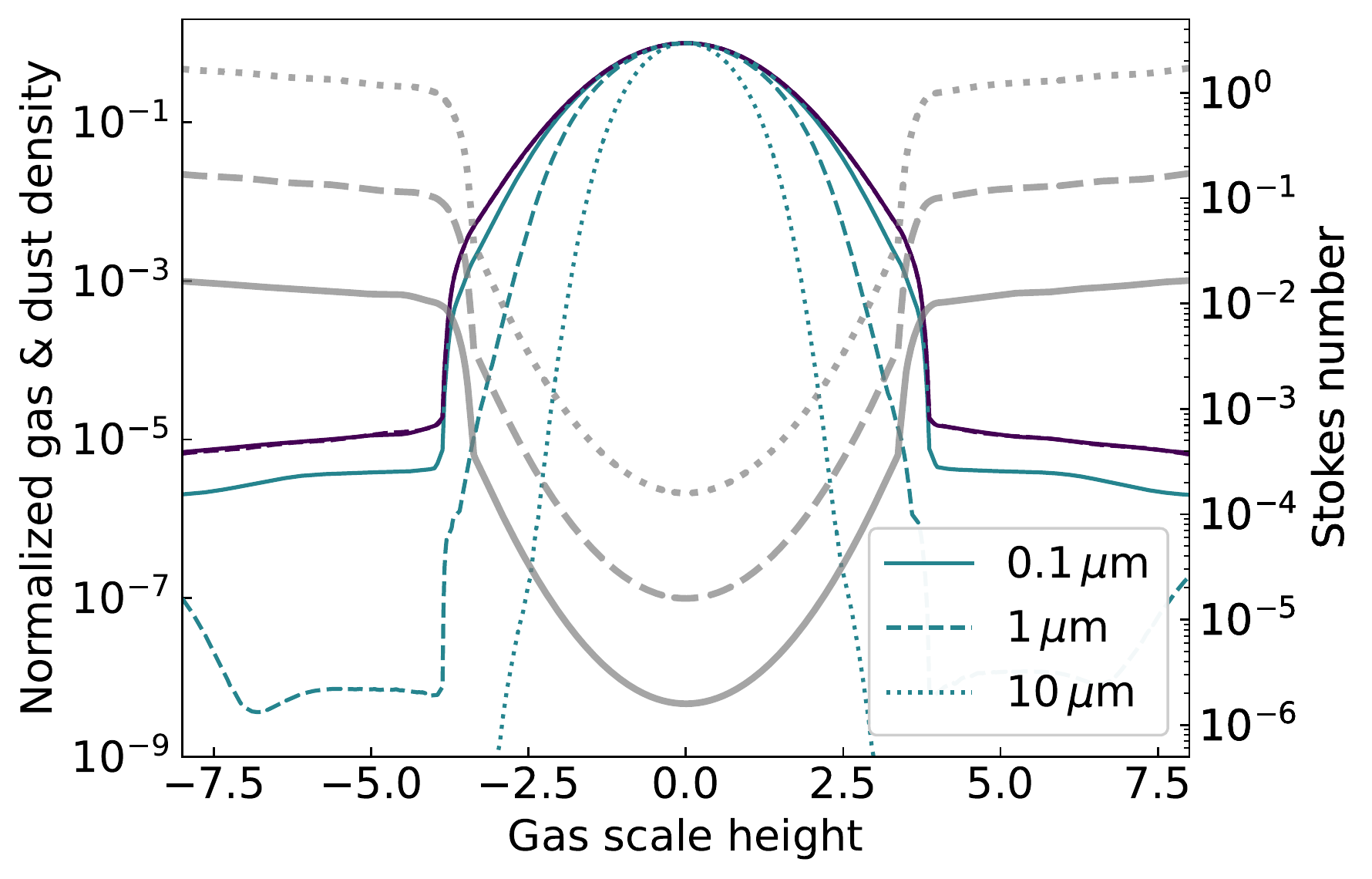}
    \caption[]{
    Normalized vertical density slices and Stokes number profiles of the \texttt{phc2}
    similar to Fig.~\ref{fig:slices_radius}
    for the grain sizes $\SI{0.1}{\micro\m}$ (\textit{straight lines}), $\SI{1}{\micro\m}$ (\textit{dashed lines}), $\SI{10}{\micro\m}$ (\textit{dotted lines})
    at 5 au.
    }
    \label{fig:slices_fluid}
\end{figure}
\begin{figure}[h]
    \centering
    \includegraphics[width=0.5\textwidth]{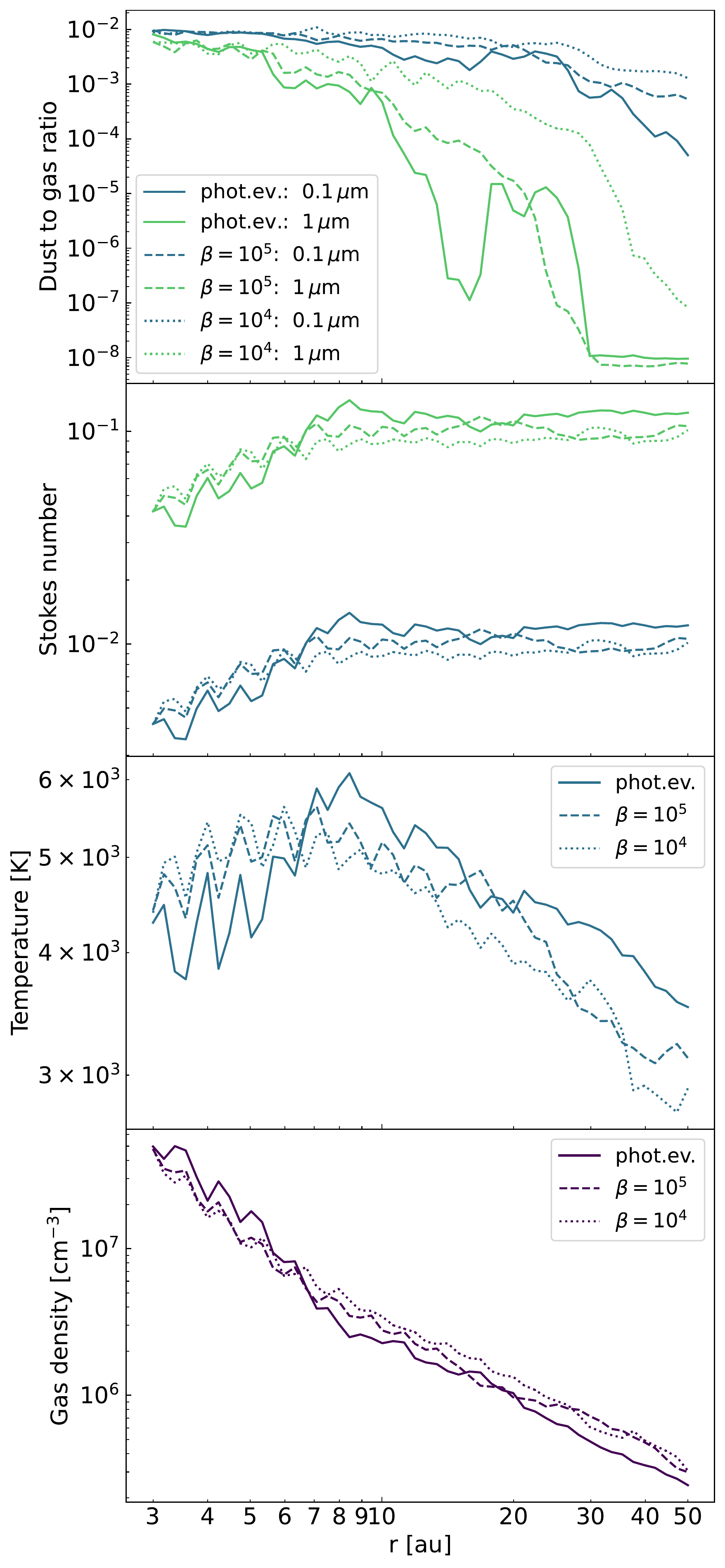}
    \caption[]{
        Physical quantities at the wind launching surface of the simulation
        runs \texttt{phc2} (\textit{straight lines}), \texttt{b5c2} (\textit{dashed lines}) and \texttt{b4c2} (\textit{dotted lines}).
        In the top two panels the blue lines represent $\SI{0.1}{\micro\m}$-sized grains,
        whereas the green lines denote $\SI{1}{\micro\m}$-sized grains.
    }
    \label{fig:launching_surface}
\end{figure}
\begin{figure*}[h]
    \centering
    \includegraphics[width=\textwidth]{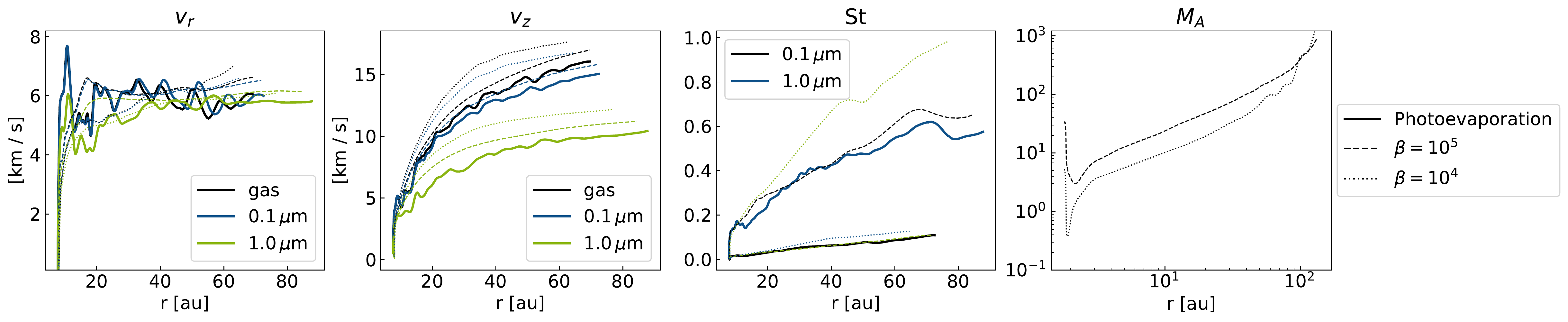}
    \caption[]{
        Streamlines starting from $r = \SI{8}{\AU}, z = \SI{2.5}{\AU}$.
        Shown are poloidal velocities of gas and dust along the streamline (\textit{left two panels})
        and the respective Stokes numbers $\mathrm{St}$ as well as the Alfv\'en Mach number $M_\mathrm{A}$ (\textit{right two panels}).
    }
    \label{fig:trajectory_analysis}
\end{figure*}
\subsection{Wind launching surface} \label{sec:launching_surface}
Since the dust entrainment efficiency seems to be dependent on the
dust reservoir at the location of the foot point of the wind, we 
therefore provide a more detailed analysis of 
the radial dependencies of the dust entrainment efficiency in this subsection. \\
We only focus on winds including photoevaporation where
the wind launching surface is marked by a sharp transition in the gas density.
Fig.~\ref{fig:slices_radius} and Fig.~\ref{fig:slices_fluid} 
visualize vertical slices of normalized gas and dust densities as well as 
the respective Stokes number profile.
As expected, in Fig.~\ref{fig:slices_radius} the dust scale height of $\SI{10}{\micro\m}$
sized grains narrows with increasing radius as the Stokes number within
the bulk of the disk increases due to the negative radial density slope. 
The kink at roughly 4.5 gas scale heights represents the wind launching surface.
Clearly, the dust scale height is too small to lift a considerable amount of 
grains to the launching surface, especially at larger radii.\\
A comparison between the different dust species at $\SI{5}{\AU}$ is given in
Fig.~\ref{fig:slices_fluid}. The trend of larger dust scale heights with
smaller Stokes numbers is readily visible. Weaker wind flows by several orders
of magnitude appear for $\SI{1}{\micro\m}$ sized grains compared to
$\SI{0.1}{\micro\m}$ particles. Only the vertical distribution of the smallest
$\SI{0.1}{\micro\m}$ grains that are well coupled to the gas appear to
correspond the gas density distribution.\\
In the following, we aim to examine the dust entrainment efficiency depending
on the radial location in the disk. We therefore traced various quantities, such as
dust-to-gas ratio, Stokes number, temperature and gas number density along the 
wind launching surface. Numerically we defined this surface to be located 
about 0.3 gas scale heights above the kink in the vertical gas density profile
in order to decrease the impact of fluctuations close to the surface of the disk.
The kink location was determined by numerically evaluating local extrema in the
second derivative of the vertical gas density profile which we find to be a robust 
method to find the launching surface. \\
The corresponding quantities at these positions 
are displayed in Fig.~\ref{fig:launching_surface} where thermally and magnetically driven 
winds with $\SI{0.1}{\micro\m}$ and $\SI{1}{\micro\m}$-sized grains are taken into account.
Clearly, the dust entrainment efficiency, defined by the dust to gas ratio close
to the wind launching front shown in the upper most panel of
Fig.~\ref{fig:launching_surface}, decreases significantly with increasing
radius.
At the inner part of the flow close to the cavity at $\SI{2}{\AU}$, $\SI{0.1}{\micro\m}$-
sized grains are basically perfectly entrained with the wind for both photoevaporative
and magnetically driven outflows since the initial dust-to-gas ratio is
$\epsilon_\mathrm{dg} = 10^{-2}$. In the outer part of the disk the dust-to-gas ratio in the wind
decreased by one to two orders of magnitude. For $\SI{1}{\micro\m}$-sized grains
the effect is more pronounced with a decline of six orders of magnitude.
It becomes apparent that generally the magnetically driven wind with $\beta = 10^4$,
represented by the dotted lines, leads to a higher dust-to-gas ratio in the wind flow
compared to weaker field strengths or a purely thermally driven wind.\\
One could argue that dust entrainment becomes less effective at larger radii
because of the more diluted gas in the wind and thereby larger Stokes numbers.
This is however not the case as shown in the second panel of Fig.~\ref{fig:launching_surface}.
In the inner part of the disk up to $\approx \SI{8}{\AU}$ to $\SI{9}{\AU}$ the Stokes number close
to the launching surface increases until it stagnates and remains constant at the
outer part of the disk. The decrease in dust entrainment efficiency can thus not 
be attributed to a weaker coupling between gas and dust in these regions.
The gas density in the wind close to the launching surface indeed decreases with
radius as shown in the bottom panel of Fig.~\ref{fig:launching_surface}.
On the other hand, the gas temperature heated by the ionizing radiation from the 
central star in turn also decreases with radius. The peak in temperature is reached 
at $\approx \SI{8}{\AU}$ to $\SI{9}{\AU}$, agreeing with the stagnation point of the Stokes number 
profile. 
By linear fitting the quantities at the wind launching surface in logarithmic space
we determine the power law slopes. 
Beyond $\SI{8}{\AU}$ the thermal velocity $v_\mathrm{th}$ decreases $\propto r^{-0.13}$
whereas the gas density profile follows the power law $\propto r^{-1.41}$. 
Looking at the definition of the Stokes number in Eq.~\ref{eq:stokes_number},
we expect an approximately constant relation $\mathrm{St} \propto \Omega_\mathrm{K} / 
(\rho_\mathrm{g} \cdot v_\mathrm{th}) \propto r^{0.04}$ with 
a Keplerian angular frequency following $\Omega_\mathrm{K} \propto r^{-1.5}$.
\subsection{Gas and dust streamlines} \label{sec:streamlines}
In Fig.~\ref{fig:trajectory_analysis} velocities, Stokes number and Alfv\'en 
Mach number are plotted for a representative streamline starting at 
$r = \SI{8}{\AU}$. Since the flow is close to a steady state we can 
consider the streamline as a proxy for the actual trajectory of a dust and gas
parcel moving in the wind.
The radial velocities in the left-most panel are almost constant with the cylindrical radius
whereas the vertical velocities in z-direction significantly increase up 
to values on the order of $\SI{15}{\km\per\second}$.
The models including magnetic effects generally lead to slightly 
faster wind flows increasing with the magnetic field strength.
Depending on the dust grain size the velocity can be significantly less
compared to the gas velocity, differing by roughly $\SI{5}{\km\per\second}$.
Dust velocities are also consistently faster in the magnetic wind case.
Looking at the Stokes numbers of the two smallest dust species,
the values up to unity are reached for micron-sized grains far from the disk.
The right-most panel demonstrates that the wind is only subalfv\'enic
for a small part close to the wind launching front. Even in the simulation 
run \texttt{b4c2} with an initial $\beta = 10^4$ the wind quickly reaches 
superalfv\'enic speeds which in consequence represents a small magnetic lever arm.
This picture is in line with recent models of magnetothermal winds 
such as \cite{Gressel2015b, bai_2017, rodenkirch_2020_wind}.
\subsection{Synthetic observations} \label{sec:synthetic_observations}
\begin{figure}[h]
    \centering
    \includegraphics[width=0.45\textwidth]{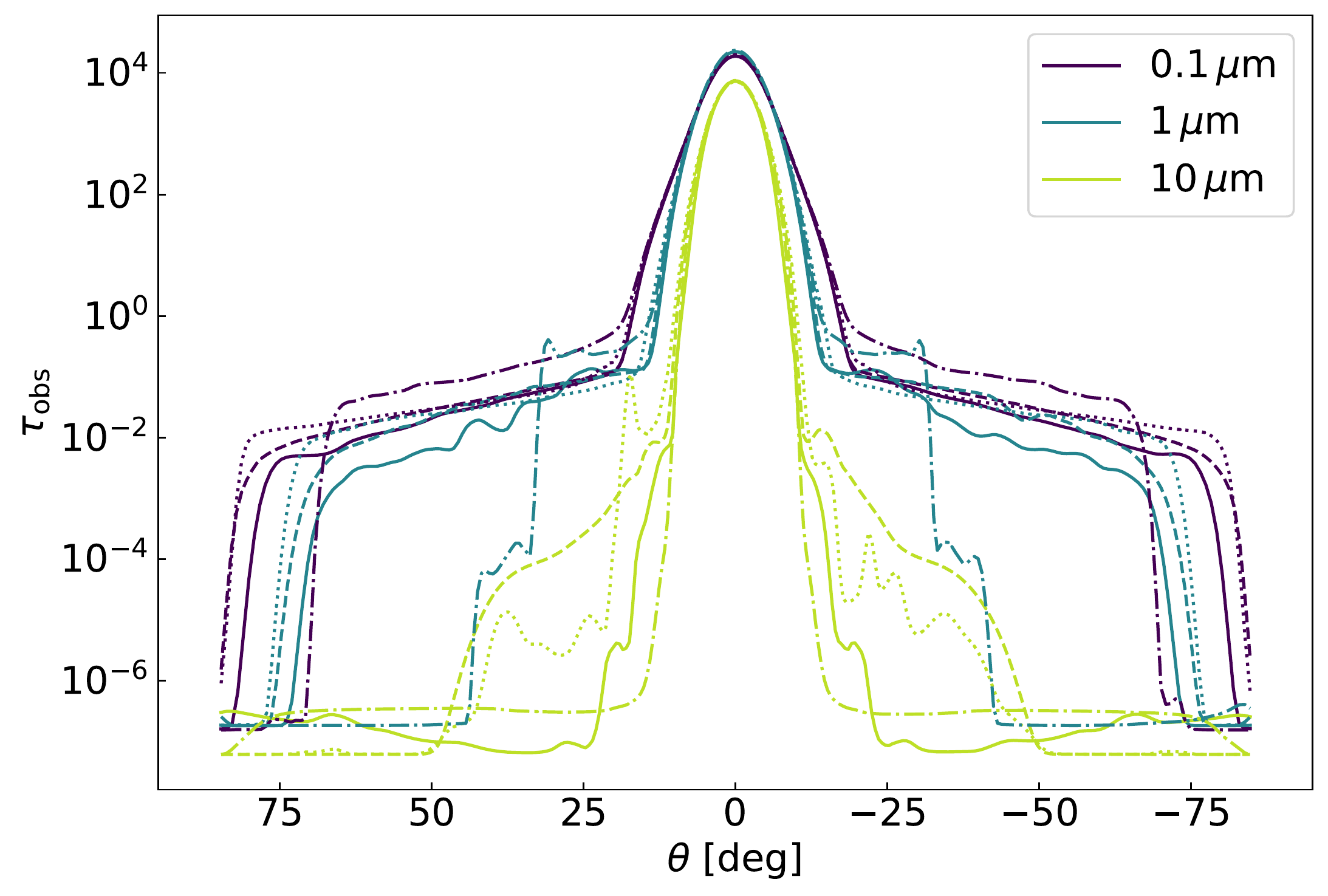}
    \caption[]{
        H-band radial optical depth $\tau_\mathrm{obs}$ originating from the central star depending \revised{on the inclination expressed in latitude}
        for three dust species with sizes $\SI{0.1}{\micro\m}$, $\SI{1}{\micro\m}$
        and $\SI{10}{\micro\m}$ indicated by the color scheme.
        The continuous lines refer to the photoevaporation simulation \texttt{phc2},
        the dashed lines to the magnetic wind \texttt{b5c2} with $\beta = 10^5$
        and the dotted lines to the run \texttt{b4c2} with $\beta = 10^4$.
    }
    \label{fig:optical_depth}
\end{figure}
\begin{figure*}
    \centering
    \includegraphics[width=\textwidth]{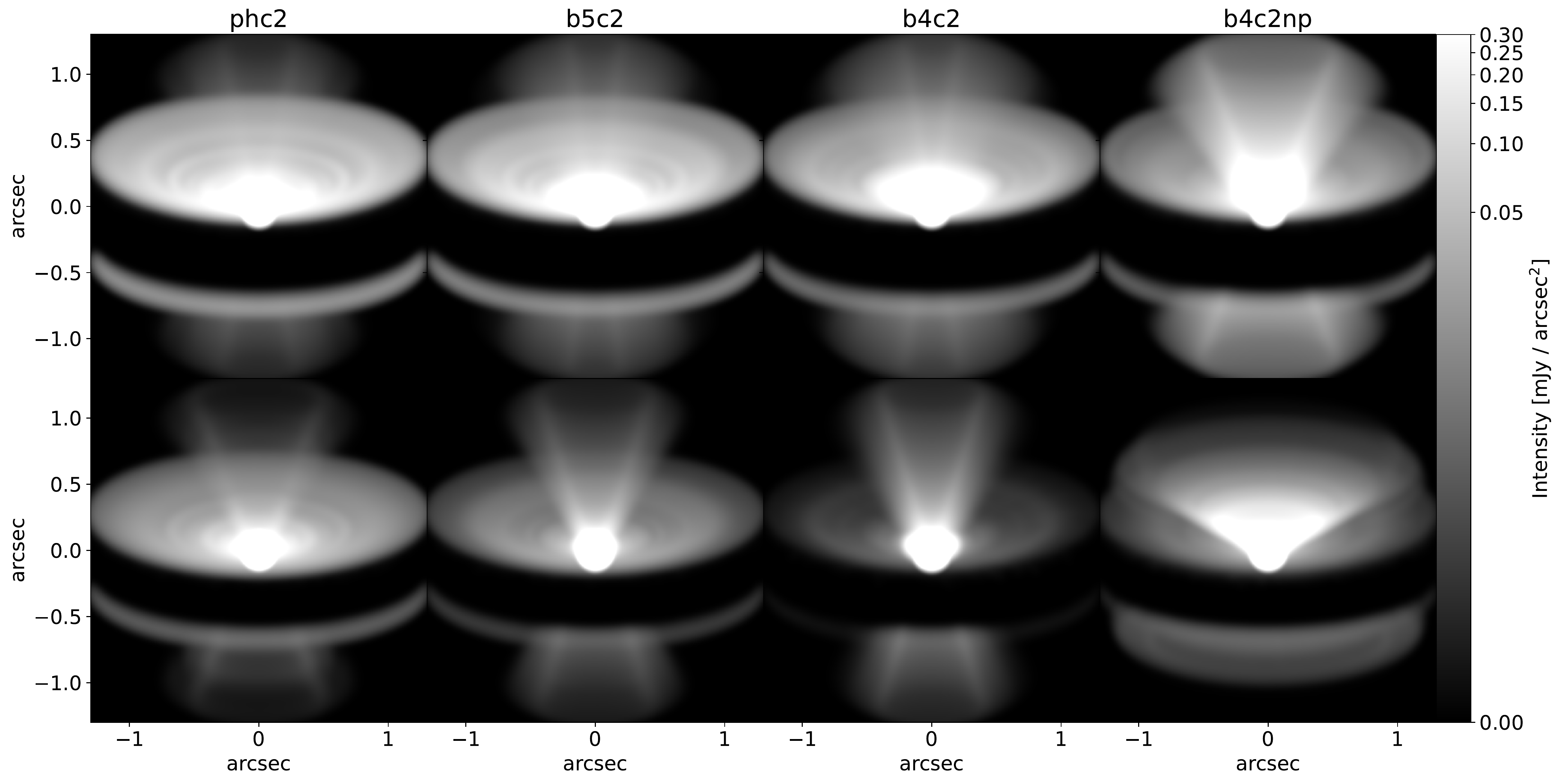}
    \caption[]{
        Radiative transfer images \revised{in the H-Band ($\lambda = \SI{1.6}{\micro\m}$)} of simulation runs with photoevaporation (\textit{left most column}), $\beta = 10^5$ (\textit{center left column}), $\beta = 10^4$ (\textit{center right column}) and a cold magnetic wind with $\beta = 10^4$ (\textit{right most column}).
        The upper panels are based on dust grains with $a = \SI{0.1}{\micro\m}$
        whereas the lower panels demonstrate synthetic images with $a = \SI{1}{\micro\m}$.
        \revised{The object is modeled at a distance of 150 pc with a solar type star at an inclination
        of \SI{20}{\degree}}.
    }
    \label{fig:synthobs_70}
\end{figure*}
\begin{figure*}
    \centering
    \includegraphics[width=\textwidth]{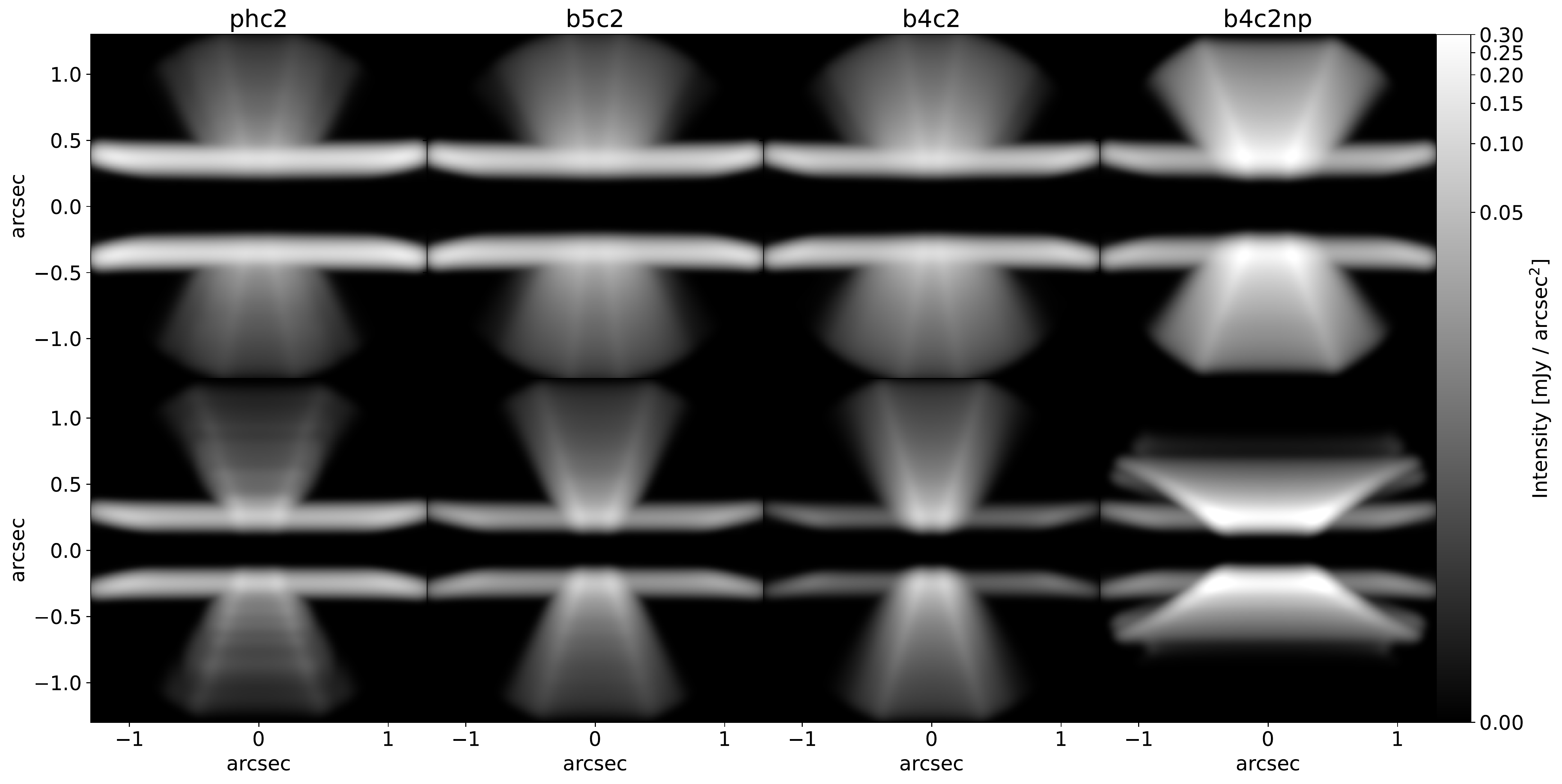}
    \caption[]{
        Radiative transfer images \revised{in the H-Band ($\lambda = \SI{1.6}{\micro\m}$)} of the same simulation runs as in Fig.~\ref{fig:synthobs_70}
        but with an \revised{edge-on view at an inclination of $\SI{0}{\degree}$.}
        }
    \label{fig:synthobs_90}
\end{figure*}
Given that the wind flow is rather thin and dust is not perfectly entrained 
for larger grain sizes and distances from the central star, the observability 
of such dusty winds deserves a closer look. 
As a first simple estimate we computed the radial optical depth depending 
on the \revised{latitude} $\theta$. The results are displayed in Fig.~\ref{fig:optical_depth}
for three dust species ranging from $\SI{0.1}{\micro\m}$ to $\SI{10}{\micro\m}$
in the H-band.
The optical depth $\tau_\mathrm{obs, i}$ of dust species i was calculated by
\begin{equation}
    \tau_\mathrm{obs, i} = \int_{r_\mathrm{min}}^{r_\mathrm{max}}{\kappa_\mathrm{sca, i} \, \rho_\mathrm{d, i}}
\end{equation} 
with the scattering opacity $\kappa_\mathrm{sca, i}$. The inner and outer radial 
boundaries are denoted by $r_\mathrm{min}$ and $r_\mathrm{max}$, respectively.
In Fig.~\ref{fig:optical_depth} the bulk of the disk appears to be optically thick
for all dust species in the H-band. The wind itself however is optically thin 
throughout the whole flow and its optical depth reaches values of 
$\tau_\mathrm{obs} \approx 10^{-1}$ in the direction of the wind base.
Naturally, the optical depth decreases with larger viewing angles due to 
the thinner dust and gas density in these regions. \\
For $\SI{1}{\micro\m}$-sized
grains the optical depth is comparable to the one for $\SI{0.1}{\micro\m}$ grains
but tapers off at a flatter angle of $\SI{70}{\degree}$ due to the 
less inclined dust flow.
Finally, the larger $\SI{10}{\micro\m}$ grains are almost irrelevant
since dust entrainment is negligible and the densities are therefore
approaching the dust density floor value.
In Fig.~\ref{fig:synthobs_70} and Fig.~\ref{fig:synthobs_90}
synthetic observations are shown with \revised{inclination angles of $\SI{20}{\degree}$
and $\SI{0}{\degree}$ in latitude, respectively}.
In comparison to the disk brightness the wind structure can be clearly 
identified as an hourglass-shaped signature.
Smaller $\SI{0.1}{\micro\m}$ grains are confined to a narrower cone
with diffuse emission towards the outer regions. A rather sharp
transition between the region devoid of dust close to the rotation axis
and the inner most part of the dusty wind is visible.\\
The wind is not completely illuminated by the stellar light because 
of the decreasing dust content towards the outer regions of the disk.
\revised{
Comparing the morphology of the 
scattered light signatures from $\SI{0.1}{\micro\m}$ grains to emission of 
$\SI{1}{\micro\m}$ grains, the larger dust particles lead to 
a conical shape with a wider opening angle with respect to the rotation axis.} This feature is 
expected from Fig.~\ref{fig:dust_inclination} where the
increased grain size leads to weaker gas-dust coupling and 
therefore causes a dust flow with a smaller inclination.
In addition, larger grains are entrained less efficiently 
in the outer regions of the disk compared to smaller grains,
due to the stronger dust settling towards the disk mid plane.
In combination with the larger opening angle of the flow
we therefore detect a less diffuse, conical shape with a sharp
transition in brightness towards the outer regions of the disk.\\
It is also visible that magnetic winds with stronger fields
confine the dust flow more towards the rotation axis and
exhibit a larger inclination angle of the dust flow.
The cold MHD wind model exhibits a brighter wind region
compared to the models including photoevaporation.
As expected, the opening angle of the $\SI{0.1}{\micro\m}$ wind flow is larger compared
to the warm wind models and resembles the emission of $\SI{1}{\micro\m}$
if photoevaporative heating is included.
In the cold wind model $\SI{1}{\micro\m}$-sized grains are 
entrained in a shallow \revised{inclination} angle and could be identified as a disk 
with a larger aspect ratio than the thermodynamics would 
otherwise allow in the system.
At a viewing angle of $\SI{20}{\degree}$ the backside of the disk almost
shadows the dusty wind, making it difficult to identify such a dusty outflow 
on the basis of a low resolution or a low signal-to-noise ratio.
Generally, we find the brightness contrast between wind and disk 
to be less in the case of $\SI{1}{\micro\m}$ grains if the viewing angle
is $\SI{20}{\degree}$.
In the edge-on view shown in Fig.~\ref{fig:synthobs_90} the wind
signature brightness is comparable to the disk brightness for 
both grain sizes.

\section{Discussion} \label{sec:discussion}
In the following we discuss limitations of our wind model
and compare the key results with other works in the literature.
An advantage of the fluid description over a Lagrangian particle formulation of
the dust dynamics in the simulations, is the ability to resolve low density regions if
a large density contrast between the midplane and the corona as well as the wind is present, which is
the case 
in the disk-wind system shown here. The disadvantage is 
the lack of proper modeling of a dust grain size distribution.
A single-size dust population is unlikely due to dust growth 
and fragmentation. The synthetic observations thus only serve as 
a hint towards the detectability of dusty winds and the expected 
shape of the flow assuming the corresponding predominant dust size.
Including multiple dust fluids in the synthetic observations 
would lead to misleading structures in the emission because of the 
rather sharp transition between the inner dust free cone and the wind region.\\
Deducing the maximum dust grain size from the wind flow angle 
seems to be rather difficult since the underlying parameter space
is degenerate. On the one hand, the opening angle of the dusty wind
increases with the grain size. On the other hand, the same effect 
occurs with colder magnetically driven winds. A distinction 
could be made by probing the wind speed, which is smaller in the cold
wind case, albeit being a specific example.\\
If the wind is optically thin for the ionizing radiation
of the central star, the difference between photoevaporative winds
and warm magnetically driven winds is minor.
In future works, the parameter space could be extended by 
varying the luminosity of the central star and by using 
a more sophisticated photoevaporation model as presented in \cite{ercolano_2021, picogna_2021}. \\
Within the simulated time frame we do not observe any significant asymmetry 
between the upper and lower wind. Such asymmetric flows were observed 
in simulations by \cite{Bethune2017a}, \cite{suriano_wind_ad_2018}
and \cite{Riols2018}. Ring formation by magnetically driven winds
does not occur in our simulation runs compared to \cite{suriano_wind_ad_2018}
and \cite{Riols2018} where ambipolar diffusion is prescribed to be the dominant
nonideal MHD effect. In their work however, larger field strength 
in the regime of $\beta = 10^3$ were probed. \cite{Riols2018} also observed
ring formation for $\beta = 10^4$ and included fluid dust focusing 
on the dust dynamics in the mid plane. \\  
Given that the same photoevaporation recipe is used in this paper \revised{as} the 
one in \cite{franz_2020} our results are compatible in this regime.
They conclude a maximum entrainable grain size of $\SI{11}{\micro\m}$.
As shown in Fig.~\ref{fig:dust_trajectories_semi} our photoevaporative 
models allow a short-term lifting of $\SI{10}{\micro\m}$ dust grains which
eventually fall back onto the disk. In the complete dust fluid picture
the small dust grain scale height prevents any dust entrainment
starting from the wind launching front. More recently,
\cite{franz_2021} studied the observability of dust entrainment
in photoevaporative winds based on their models in \cite{franz_2020}.
They conclude that the wind signature in scattered light is 
too faint to be observed with SPHERE IRDIS, but might be detectable 
with the JWST NIRCam under optical conditions.
In their synthetic observations a similar \enquote{chimney}-shaped
wind emission is visible compared to our photoevaporative models. 
\\
\cite{giacalone_2019} found a positive correlation between
gas temperature and maximum dust grain size in their semi-analytical
magneto-centrifugal wind model which corresponds best to our cold magnetically
driven wind model \texttt{b4c2np}. Our results corroborate their 
findings of a maximum entrainable dust size in the submicron regime
in a cold wind around a typical T Tauri star.
Our models including photoevaporation furthermore agree
with \cite{hutchison_2016b} where the maximum grain size was determined 
to be $< \SI{10}{\micro\m}$ depending on 
the disk radius for a typical T Tauri star. They additionally 
argue that dust settling most certainly limits this maximum size further
which we confirm in our models, where the diminishing dust entrainment 
efficiency can be attributed to the dust settling towards the mid plane.\\
In recent scattered light observations of RY Tau \citep{garufi_2019}
a broad dusty outflow obstructing the underlying disk was detected. 
Both lobes are inclined by roughly $\SI{45}{\degree}$.
The presence of a jet in this system advocates a magnetic wind
launching mechanism that causes the features observed in the near-infrared.
We do not observe an optically thick wind that would be able to completely obstruct the disk, 
since the dust densities are too low, especially
in the models including photoevaporation. A possibility might hence
be a slow, cold, magnetic wind at higher field strengths. 
The large opening angle in the observation would suggest a wind flow
containing dust grains of the size of several $\SI{}{\micro\m}$
if photoevaporation would be the significant wind mechanism. 
If a cold magnetically driven wind was the dominant factor, 
smaller submicron grains would be the solution compatible with our
models. \\
In this work we neglected the effect of radiation pressure
that may be able to blow out small dust grains. \cite{franz_2020}
argued that a photoevaporative wind can entrain grains of sizes roughly 20 times larger compared 
to a compatible radiation pressure model from \cite{owen_kollmeier_2019}.
Since we use similar parameters
for the photoevaporation model, the effect of radiation pressure should
not significantly affect the dust dynamics in the studied parameter space.\\
In the photoevaporation models the gaseous part of the wind reaches temperatures
of up to $10^4\,\mathrm{K}$. We verified that the dust temperature 
stays below the dust sublimation temperature in Appendix~\ref{sec:dust_appendix}.

\section{Conclusion}
We presented a suite of fully dynamic, global, multifluid simulations
that model dust entrainment in photoevaporative as well as magnetically 
driven disk wind using the FARGO3D code. 
We demonstrated that both types of winds are able to efficiently 
transport dust in the wind flow and addressed the observability of 
such dusty winds with the help of the RADMC-3D radiative transfer code.
In the following we summarize the main results:
\begin{itemize}
\item The maximum entrainable dust grain size $a_\mathrm{max}$ both depends
on the type of the wind launching mechanism and the turbulent 
diffusivity of the disk. Whereas magnetic winds including XEUV-photoevaporation
only show minor differences in $a_\mathrm{max}$, ranging from $\SI{3}{\micro\m}$
to $\SI{6}{\micro\m}$, cold magnetically driven winds at comparable field
strengths only entrain submicron sized grains efficiently.
\item With increasing radius the dust to gas ratio in the wind drops rapidly, 
mainly due to the smaller dust scale height in comparison to the gas scale height.
Dust grains are unable to reach the wind launching surface in the outer regions of 
the disk and the dust content in the wind starting from these locations hence decreases.
In the case of warm, ionized winds the Stokes number stays approximately 
constant with radius beyond $\SI{8}{\AU}$. Generally, magnetically driven winds
including photoevaporation are slightly more efficient in entraining dust compared 
to pure photoevaporative flows.
\item The dust flow angle $\theta_\mathrm{d}$ decreases with 
increasing grain size to the weaker coupling between gas and dust.
In warm, ionized winds the flow angle reaches values of $\SI{67}{\degree} \leq \theta_\mathrm{d} \leq \SI{69}{\degree}$
for $a = \SI{1}{\micro\m}$ sized grains.
At a grain size of $a = \SI{3}{\micro\m}$ a photoevaporative dust flow 
provides $\theta_\mathrm{d} \approx \SI{36}{\degree}$ whereas 
a warm magnetic wind leads to a steeper dust flow with $\theta_\mathrm{d} \approx \SI{45}{\degree}$.
In a cold magnetic wind the flow angle is significantly smaller, 
being $\theta_\mathrm{d} \approx \SI{59}{\degree}$ for $a = \SI{0.1}{\micro\m}$
and $\theta_\mathrm{d} \approx \SI{25}{\degree}$ for micron-sized grains.
\item Radiative transfer images in the H-band suggest a rather weak, 
optically thin emission for photoevaporative winds. A more pronounced, conical 
shape is visible considering $\SI{1}{\micro\m}$ grains. The larger dust flow 
angle $\theta_\mathrm{d}$ leads to a narrower emission region 
in the case of warm, ionized magnetic winds. Cold magnetic winds
appear with a significantly larger opening angle with comparable grain sizes. 
The emission region of the wind appears brighter and more easily detectable
due to the increased dust density of the slower magnetic wind.
\end{itemize}
Our results presented in this paper might help to constrain the 
magnetic field strength as well as the dominating wind mechanism 
if such dust signatures will be detected in future observations.
The dust grain size could be determined via the inclination angle 
of the flow. Although the parameter space is degenerate, probing 
the wind speeds might help to differentiate between thermally and 
cold magnetically driven winds. 

\label{sec:conclusion}
\begin{acknowledgements}
  We would like to thank R. Franz for helpful discussions.
  We furthermore thank the anonymous referee for 
  helpful comments that further improved the manuscript.
  Authors Rodenkirch and Dullemond acknowledge funding from
  the Deutsche Forschungsgemeinschaft (DFG, German Research Foundation) - 325594231 under grant DU 414/23-1.
  The authors acknowledge support by the
  High Performance and Cloud Computing Group at the Zentrum f\"ur
  Datenverarbeitung of the University of T\"ubingen, the state of
  Baden-W\"urttemberg through bwHPC and the German Research Foundation (DFG)
  through grant INST\,37/935-1\,FUGG.
  Plots in this paper were made with the Python library \texttt{matplotlib} \citep{hunter-2007}.
\end{acknowledgements}

\bibliographystyle{aa}
\bibliography{main}

\begin{appendix}
\section{Dust temperature estimate} \label{sec:dust_appendix}
The relatively high temperatures in the ionized wind models heat 
the dust grains that are entrained in the gas flow. We show 
in the following that dust particles do not reach the dust sublimation
threshold and would thus remain solid along their trajectories.
The expression of the heating rate per unit surface $q_\mathrm{d}$ is described in \cite{gombosi_1986},
taking into account both contact and friction heating. 
We follow the approach of \cite{stammler_2014} in this regard and evaluate 
the subsequent expressions:
\begin{equation}
    q_\mathrm{d} = \rho_\mathrm{g} C_\mathrm{H, d} \left(T_\mathrm{rec} - T_\mathrm{d}\right) |v_\mathrm{g} - v_\mathrm{d}| \,,
\end{equation}
where the heat transfer coefficient $C_\mathrm{H, d}$ is expressed as
\begin{equation}
C_\mathrm{H, d} = \frac{\gamma + 1}{\gamma - 1} \frac{k_\mathrm{B}}{8 \mu m_\mathrm{p} s^2} \left[\frac{s}{\sqrt{\pi}} \mathrm{exp}\left(-s^2\right) + \left(\frac{1}{2} + s^2\right) \mathrm{erf}\left(s\right)\right] \,,
\end{equation}
with the error function $\mathrm{erf}$ and the normalized difference between dust and gas velocities:
\begin{equation}
s = \frac{|v_\mathrm{g} - v_\mathrm{d}|}{\sqrt{2 k_\mathrm{B} T_\mathrm{g} / (\mu m_\mathrm{p})}} \,.
\end{equation}
The recovery temperature $T_\mathrm{rec}$ is given by the expression \citep{gombosi_1986}:
\begin{equation}
    T_\mathrm{rec} = \frac{T_\mathrm{g}}{\gamma - 1} \left[ 2\gamma + 2(\gamma - 1)s^2 - \frac{\gamma - 1}{\frac{1}{2} + s^2 + \frac{s}{\sqrt{\pi}} \mathrm{exp}\left(-s^2\right) \mathrm{erf}^{-1}\left(s\right) } \right] \,.
\end{equation}
The dust equilibrium temperature is then finally computed by solving 
\begin{equation}
    q_\mathrm{d} + \frac{R_*}{r^2} \epsilon_\mathrm{d} \sigma_\mathrm{SB} T_*^4 - 4\sigma_\mathrm{SB} T_\mathrm{d}^4 = 0
\end{equation}
with the thermal cooling efficiency factor 
\begin{equation}
    \epsilon_\mathrm{d} = \frac{\kappa_\mathrm{P} (T_*)}{\kappa_\mathrm{P} (T_\mathrm{d})} \,,
\end{equation}
using the Planck mean opacity $\kappa_\mathrm{P}$ with the corresponding temperature.
\begin{figure}
    \centering
    \includegraphics[width=0.5\textwidth]{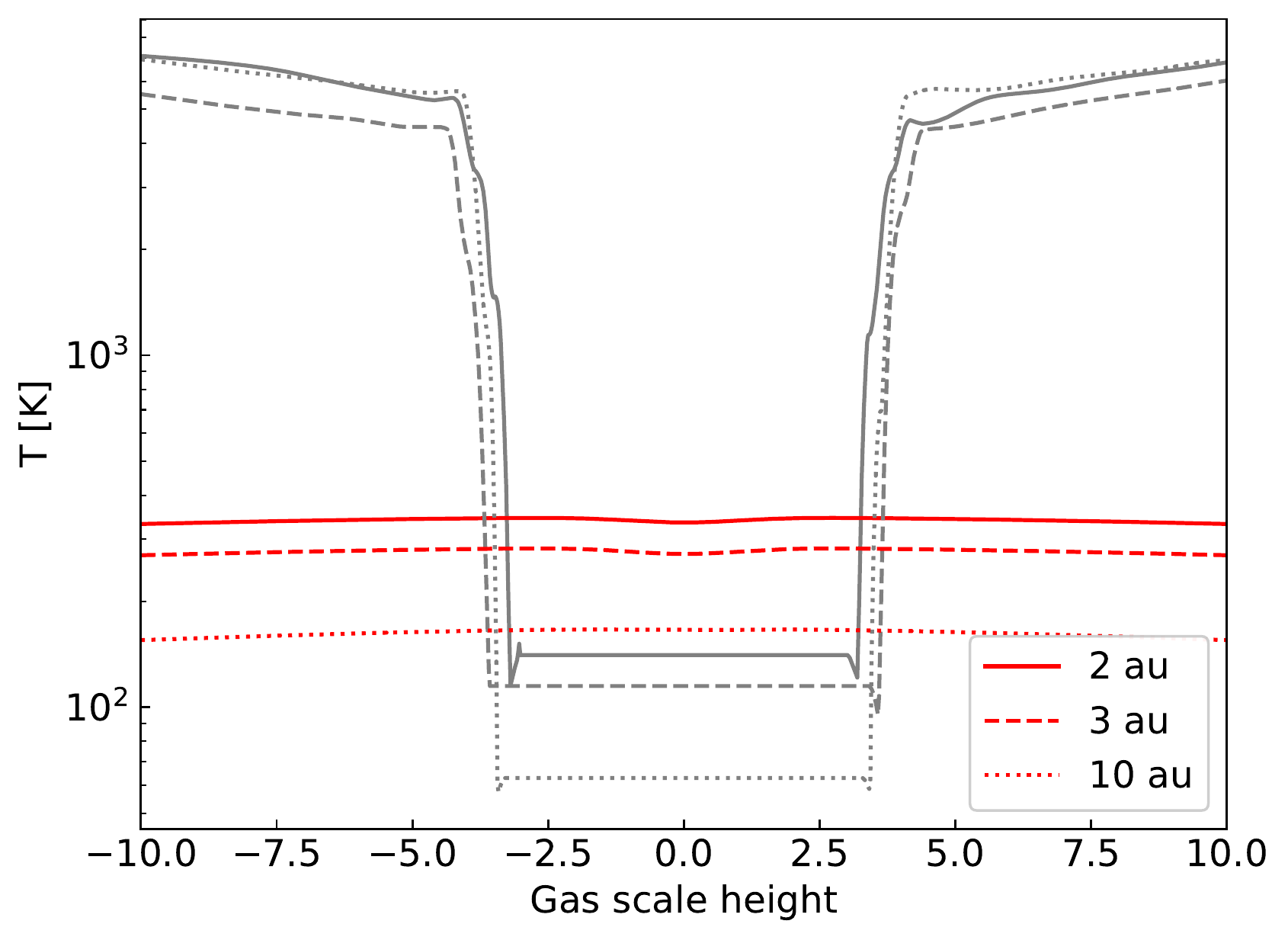}
    \caption[]{
        Vertical slices of gas and dust temperatures at different radial 
        location in the disk and wind region of the photoevaporation simulation \texttt{phc2}. 
        The gray lines represent the gas 
        temperatures whereas the red lines display the equilibrium 
        dust temperatures of the $\SI{0.1}{\micro\m}$ grains.
    }
    \label{fig:dust_temp}
\end{figure}
Fig.~\ref{fig:dust_temp} provides the solutions of the equilibrium dust temperatures
in comparison with the gas temperatures in a photoevaporative flow 
including $\SI{0.1}{\micro\m}$ grains. We assume a sun-like star in 
the computation and find maximum dust temperatures of $T_\mathrm{d} \approx \SI{345}{\kelvin}$
at a radial distance of $r = \SI{2}{\AU}$. 
Due to the thin gas in the wind region the ambient heating has an almost
negligible impact on the dust temperature. Without the flux of the central 
star the dust grains would reach $T_\mathrm{d} \approx \SI{40}{\kelvin}$
only being heated by the ionized gas.
In the wind flows presented here dust temperatures are clearly below
the dust sublimation threshold which would be $\approx \SI{980}{\kelvin}$
assuming pyroxene as the grain material embedded in a thin medium 
of the density $\rho_\mathrm{g} = \SI{1e-16}{\g\per\cm^3}$
\citep{pollack_1994}.
\section{Ionization fraction} \label{sec:ion_appendix}
\begin{figure}
    \centering
    \includegraphics[width=0.5\textwidth]{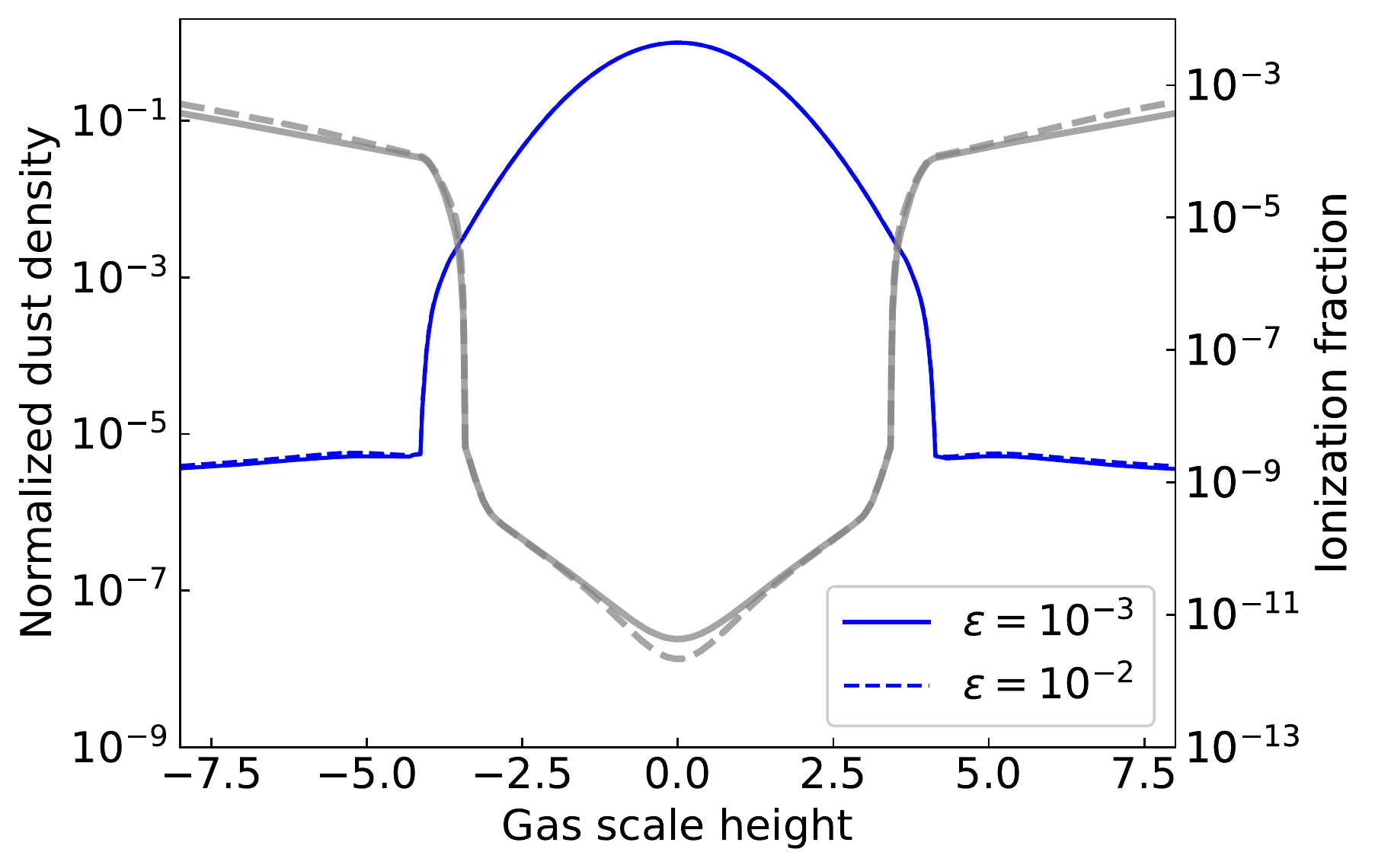} 
    \caption[]{
        Normalized dust density (\textit{blue lines}) and 
        ionization fraction (\textit{gray lines}) in the simulations
        \texttt{b4c2} and \texttt{b4c2eps2} with a dust-to-gas ratio
        of $\epsilon = 10^{-3}$ and $\epsilon = 10^{-2}$, respectively.
        The snapshot is taken after 1000 orbits at $\SI{1}{\AU}$.
    }\label{fig:ion_fraction}
\end{figure}

Since the ionization fraction depends on the dust-to-gas ratio $\epsilon_\mathrm{ion}$
in the disk a comparison between $\epsilon_\mathrm{ion} = 10^{-3}$ and $\epsilon_\mathrm{ion} = 10^{-2}$
of the simulation runs \texttt{b4c2} and \texttt{b4c2eps2} is shown 
in Fig.~\ref{fig:ion_fraction}.
The ionization is relatively insensitive to the increased dust-to-gas ratio
and only decreases by a small amount in the weakly ionized disk mid plane.
In the dust density slice the effects are negligible. 
\section{Super time stepping} \label{sec:sts_appendix}
This section describes the implementation of the Super
Time-Stepping technique (STS) for parabolic or mixed advection-diffusion
problems. Introduced by \cite{alexiades_1996}, the first-order method divides
one explicit super step $\Delta T$ in multiple substeps $\tau_1, \tau_2, \ldots
\tau_\mathrm{N}$. The algorithm ensures stability while maximizing $\Delta T =
\sum_\mathrm{i}^{\mathrm{N}} \tau_\mathrm{i}$ The optimized length of each
substep is based on Chevychev polynomials:
\begin{equation}
    \tau_\mathrm{j} = \Delta t_\mathrm{exp} \left[ (\nu - 1) \, \mathrm{cos} \left( \frac{2 j - 1}{N} \frac{\pi}{2} \right) + \nu + 1 \right] \,,
\end{equation}
where $\Delta t_\mathrm{exp}$ is the explicit time step. The sum of all
substeps then gives:
\begin{equation}
    \Delta T = \Delta t_\mathrm{exp} \frac{N}{2 \sqrt{\nu}} \left[ \frac{(1 + \sqrt{\nu})^{2 N} - (1 - \sqrt{\nu})^{2 N}}{(1 + \sqrt{\nu})^{2 N} + (1 - \sqrt{\nu})^{2 N}} \right] \,.
\end{equation}
The parameter $\nu$ adopts values between 0 and 1. In the limit of $\nu
\rightarrow 0$ the method is $N$ times faster than the explicit integration.
Lower values of $\nu$ however can lead to oscillations and eventually to an unstable solution.
In FARGO3D the magnetic field is updated with the method of characteristics and
constrained transport (MOCCT) \citep{hawley_stone_mocct} whereby the
constrained transport method described in \cite{evans_1988_ct} is used
\citep{fargo3d}. In this method the evolution of the magnetic field is
determined by computing the electromotive forces (EMFs) \cite{fargo3d}:
\begin{equation}
    \mathbf{E} = \mathbf{v} \times \mathbf{B} \,.
\end{equation} 
We compute the contribution to the total EMFs of each
substep j with duration $\tau_\mathrm{j}$. The result is added to the EMFs, 
weighted by a factor
of $\tau_\mathrm{j} / \Delta T$. The magnetic field updates at every substep
are only applied on a temporary buffer field and the main B-field is then updated
with the total EMFs, containing contributions from nonideal terms and the
ordinary MHD dynamics.\\
In order to test the correctness of the implementation, a decaying standing
Alfv\'en wave test
for ambipolar diffusion as described in \cite{choi_sts} Sec. 4.2 is used.
The time dependence of the first normal mode is given by \cite{choi_sts}:
\begin{equation}
    h(t) = h_0 \, \mathrm{sin}(\omega_R t) e^{\omega_I t}
\end{equation}
with $\omega = \omega_R + i \omega_I$.
The terms $\omega_R$ and $\omega_I$ are the real and imaginary parts of the
angular frequency of the Alfv\'en wave. The dispersion relation is given by \cite{choi_sts}:
\begin{equation}
    \omega^2 + i \frac{c_\mathrm{A}^2 k^2}{\langle \sigma v \rangle_\mathrm{i} n_\mathrm{i}} \omega + c_\mathrm{A}^2 k^2 = 0 \,,
\end{equation}
where $k$ is the real wave number and $c_\mathrm{A}$ the characteristic Alf\'en speed.\\
In the test according to \cite{choi_sts} the magnetic field is set to $\mathbf{B} = B_0 \hat{\mathrm{x}}$ with $B_0 = 1$.
The density is uniformly set to 1. The wave is initialized in the $\hat{\mathbf{x}}$ direction with an initial velocity of:
\begin{equation}
    v_\mathrm{y} = v_0 c_\mathrm{A} \, \mathrm{sin}(k x) \,.
\end{equation}
The amplitude of the perturbation is set to $v_0 = 0.1$, the characteristic Alf\'en velocity to $1 / \sqrt{2}$, the ion density 
to $\rho_\mathrm{i} = 0.1$ and the wave number to $k = 2 \pi / L$ where $L = 1$ is the domain size.
The collision rate coefficient $\langle \sigma v \rangle_\mathrm{i}$ is varied from 100 to 1000.
Instead of an oblique wave test, the simulations are run in a one-dimensional configuration.
\begin{figure}
    \centering
    \includegraphics[width=0.5\textwidth]{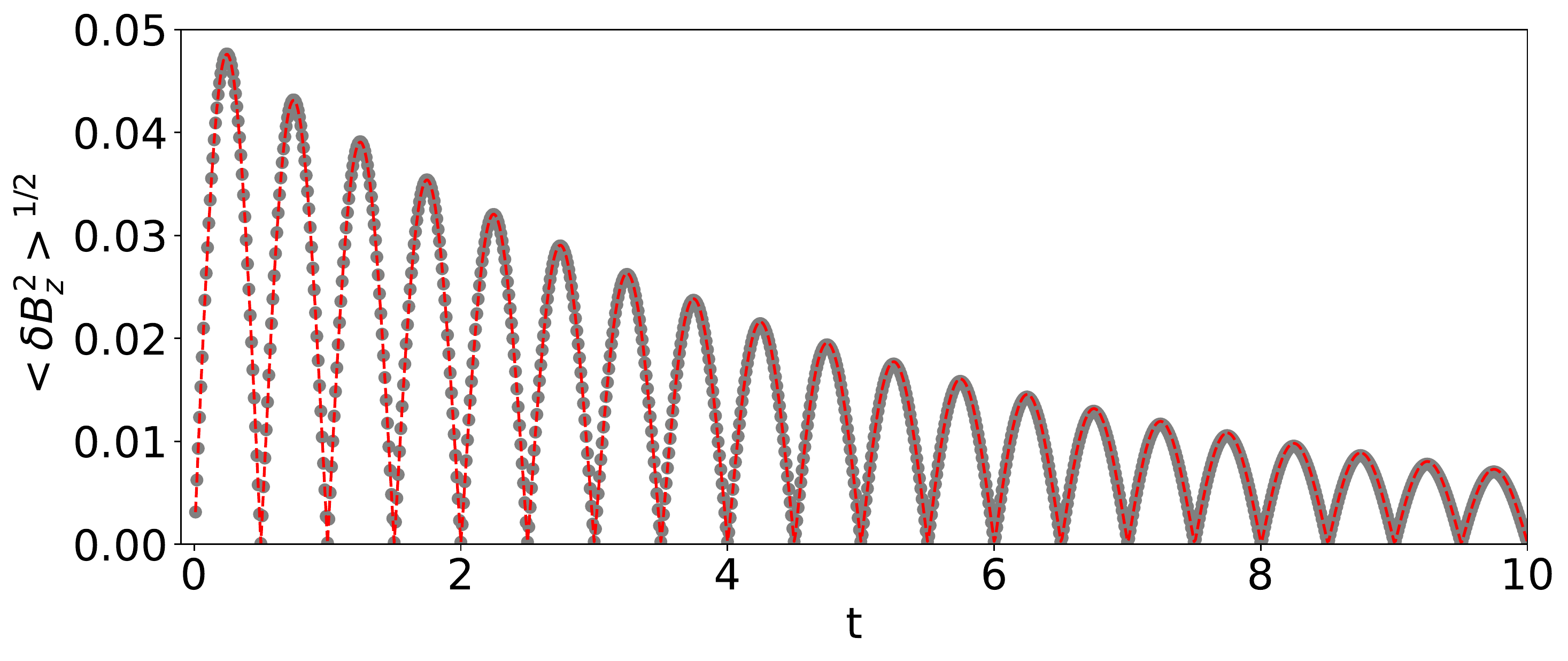} 
    \caption[Ambipolar diffusion test 1]{
        Decaying Alfv\'en wave test for $\langle \sigma v \rangle_\mathrm{i} =
        1000$. The dots correspond to the simulation results whereas the dashed
        line is the analytical solution.
    }\label{fig:am_test_1000}
\end{figure}
\begin{figure}
    \centering
    \includegraphics[width=0.5\textwidth]{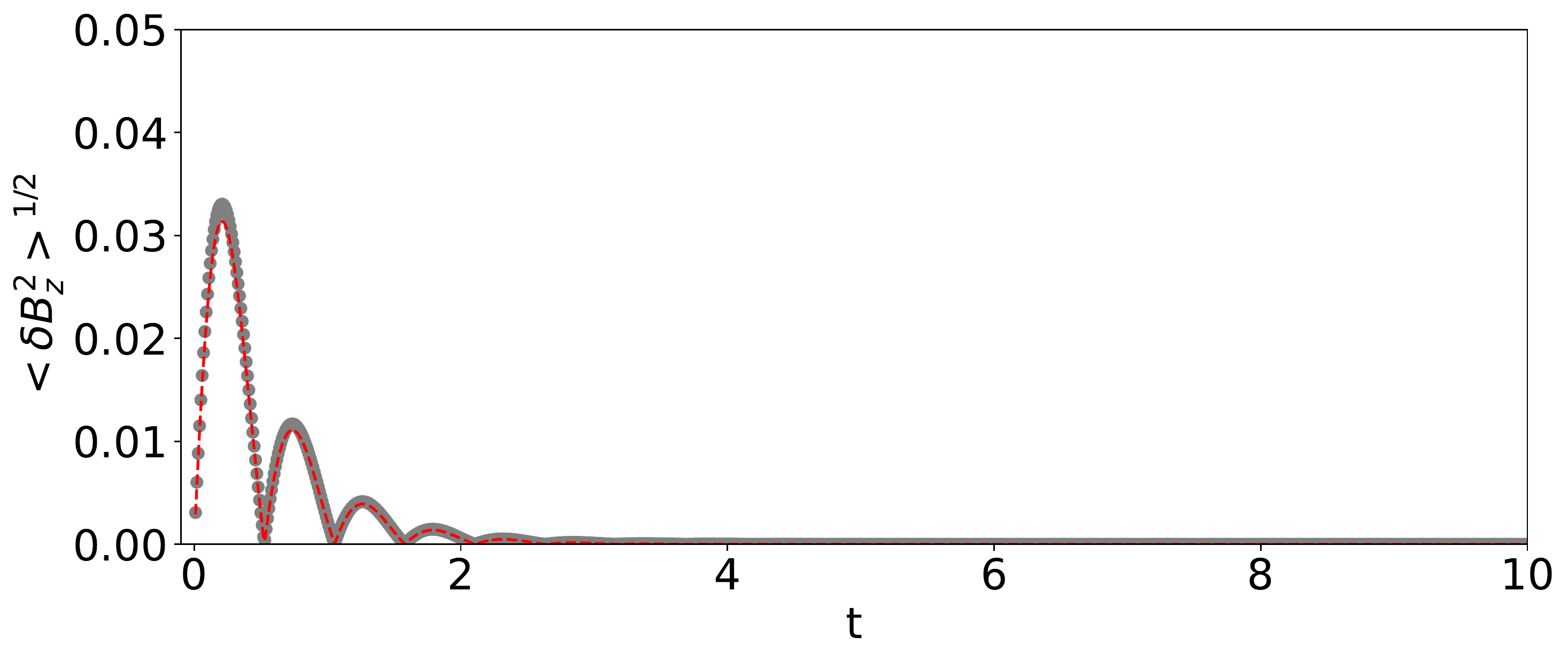} 
    \caption[Ambipolar diffusion test 1]{
        Decaying Alfv\'en wave test for $\langle \sigma v \rangle_\mathrm{i} = 100$ similar to Fig.~\ref{fig:am_test_1000}   
        }\label{fig:am_test_100}
\end{figure}
In Fig.~\ref{fig:am_test_1000} and Fig.~\ref{fig:am_test_100} the test results
are shown and the numerical results fit the analytical solution well. 
Stability is found to be
fulfilled up to five substeps which is sufficient for the simulation runs presented in this paper.
An overall loss of accuracy in comparison to
the solution without STS is also anticipated since this approach is only
first-order accurate.
\end{appendix}

\end{document}